\documentclass[10pt]{article}

\usepackage{fullpage}
\usepackage{amsmath}
\usepackage{amssymb}
\usepackage{graphicx}
\usepackage[all,poly,curve,knot,cmtip]{xy}
\usepackage{citesort}

\newcommand {\be} {\begin{eqnarray*}}
\newcommand {\ee} {\end{eqnarray*}}
\newcommand {\bea} {\begin{eqnarray}}
\newcommand {\eea} {\end{eqnarray}}
\newcommand \foot[1]{\footnotemark\footnotetext{#1}}
\newcommand {\bm}[1] {\boldsymbol{#1}}
\newcommand{\ket}[1]{| {#1} \rangle}
\newcommand{\bra}[1]{\langle {#1} |}
\newcommand{\inner}[2]{\langle {#1} | {#2} \rangle}

\newcommand{\map}[3]{{#1}:{#2}\rightarrow{#3}}
\newcommand{\poisson}[2]{\lbrace {#1} , {#2} \rbrace}

\newcommand{\fdiff}[2]{\frac{\delta{#1}}{\delta{#2}}}
\newcommand{\pdiff}[2]{\frac{\partial{#1}}{\partial{#2}}}
\newcommand{\tdiff}[2]{\frac{d{#1}}{d{#2}}}
\newcommand{\fsent}[2]{\delta{#1}/\delta{#2}}
\newcommand{\psent}[2]{\partial{#1}/\partial{#2}}
\newcommand{\tsent}[2]{d{#1}/d{#2}}

\title{Knot theory and a physical state of quantum gravity}
\author{\textbf{Tom$\acute{\mbox{a}}\check{\mbox{s}}$ Liko}${}^{\dagger}$
and \textbf{Louis H. Kauffman}${}^{\ddagger}$ \\
\\ {\small \it ${}^{\dagger}$Department of Physics, University of Waterloo} \\
{\small \it Waterloo, Ontario, Canada, N2L 3G1} \\
{\small \it Email address: tliko@uwaterloo.ca} \\
{\small \it ${}^{\ddagger}$Department of Mathematics, Statistics and Computer Science} \\
{\small \it University of Illinois at Chicago, Chicago, Illinois, USA 60607-7045} \\
{\small \it Email address: kauffman@uic.edu}}

\begin{document}

\maketitle

\begin{abstract}

We discuss the theory of knots, and describe how knot invariants arise
naturally in gravitational physics.  The focus of this review is to
delineate the relationship between knot theory and the loop representation
of non-perturbative canonical quantum general relativity (loop quantum
gravity).  This leads  naturally to a discussion of the Kodama wavefunction,
a state which is conjectured to be the ground state of the gravitational
field with positive cosmological constant.  This review can serve as a
self-contained introduction to loop quantum gravity and related areas.  Our
intent is to make the paper accessible to a wider audience that may include
topologists, knot-theorists, and other persons innocent of the physical
background to this approach to quantum gravity.

\end{abstract}

\hspace{0.3cm}\textbf{PACS}: 04.20.Fy; 04.60.Ds; 04.60.Pp

\section{Introduction}

\emph{``Physical laws should have mathematical beauty.''}  $\sim$ P A M Dirac,
Moscow 1955.

The oldest known invariant of knots is the Gauss linking number, which was
discovered by Johannes Gauss \cite{gauss,ashcor} in 1833.  Physically this
is the work done in transporting a magnetic monopole around a closed loop
$L_{1}$ through a magnetic field that is produced by a current moving through
a closed loop of wire $L_{2}$.  The electric current and magnetic field have
fixed directions, so this configuration may be thought of as an oriented link
with two components.  Topologically the invariant measures the degree to which
the two loops are linked.  Furthermore, the invariant is well defined so long
as the two loops do not touch or intersect.  Gauss calculated the work done on
the monopole to be the double line integral
\bea
W = \frac{mI}{4\pi}\oint_{L_{2}}ds\oint_{L_{1}}dt\epsilon_{ijk}
    \dot{L}_{2}^{i}(s)\dot{L}_{1}^{j}(t)
    \frac{L_{2}^{k}(s)-L_{1}^{k}(t)}{|L_{2}(s)-L_{1}(t)|^{3}}.
\eea
The invariant is then the work divided by the current $I$ and magnetic charge $m$
of the monopole.  A generalization of this invariant was given by Edward Witten
\cite{witten1} in 1989 as the expectation value of certain gauge-invariant
functionals, showing that topological invariants of links are fundamental in
gauge theories.

The analytical study of knots began in the 1880's, after James Maxwell \cite{maxwell}
had established the equations of electrodynamics on a firm mathematical foundation.
At that time quantum mechanics had not yet been postulated, and the remarkable
stability of atoms was a mystery.  It seemed quite natural therefore to explain
these structures purely in terms of Maxwell's theory.  William Thomson (Lord Kelvin)
\cite{keltai} put forward the hypothesis that atoms were knotted flux tubes of the
aether.  His motivation really came from the strong analogy between the vacuum
equations for the electric field and the fluid flow equations of hydrodynamics, but
this also gave a satisfactory explanation for the stability and large variety of
atoms.  This theory was taken seriously for a while and attracted a modest following.
In particular, Peter Tait \cite{tait} began his classification of knots as projections
onto a two-dimensional plane.  Along the way he noticed some properties which led him
to what we now call the Tait conjectures.  Most of these were not proved until the
1980's after Vaughan Jones \cite{jones} discovered a polynomial invariant which now
bares his name.  Indeed this discovery marked a revival of knot theory long after the
inevitable failure of the vortex model when the principle of relativity by Albert
Einstein \cite{einstein} did away with the aether, and with it the possibility of
vortices in the aether.  In the meantime, (up until 1984) knot theory had developed
into a vigorous branch of low dimensional topology, using all the techniques of modern
geometric and algebraic topology.  The advent of the Jones polynomial and its
generalizations brought new techniques and startling new relationships with physics
into this already active field.  After Jones' discovery, many papers began to appear
in which new link invariants were defined.  A definition of the Jones polynomial
was given by Louis Kauffman \cite{kauffman} as a state summation in analogy with
partition functions in certain low dimensional systems in statistical mechanics (such
as the Potts model).  This model is called the bracket polynomial or the Kauffman bracket.
The Jones polynomial is the product of the bracket polynomial and the self-linking number
(the writhe) of a given link.  Seminal work of Witten \cite{witten1} then showed that link
invariants could be constructed via gauge theory.  Witten accomplished this relationship
by calculating the expectation values of Wilson loops using facts from two-dimensional
conformal field theory.  Simultaneously and independent of this result, a representation
for four-dimensional canonical quantum gravity was proposed by Carlo Rovelli and Lee
Smolin \cite{rovsmo1,rovsmo2}, which exploits the topological invariance of physical
states in the so-called loop space, and are otherwise invariant under spacetime
diffeomorphisms.  This generalized earlier work of Ted Jacobson and Smolin \cite{jacsmo},
who showed that loop solutions solved all the constraints of the Ashtekar phase space.
Rovelli and Smolin were motivated by an idea due to Chris Isham \cite{ishkak}, that a
quantization of general relativity may involve a non-canonical Poisson algebra.
Remarkably, however, this loop representation is closely related to an old idea of
Roger Penrose \cite{penrose}, whereby the fundamental structure of spacetime is
combinatorial, given by the spin networks.  Soon after, many exact states satisfying
the quantum constraints were found by Hideo Kodama \cite{kodama1,kodama2}, Viqar Husain
\cite{husain} and Bernd Br$\ddot{\mbox{u}}$gmann, Rodolfo Gambini and Jorge Pullin
\cite{bgp}; these were all shown to be related to the Jones polynomial and other similar
knot invariants.  As we will see, none of the ideas presented here are independent of each
other.

The outline of the paper is as follows.  We begin $\S$II by reviewing the foundations
of non-Abelian gauge theories.  We define the holonomy of the connection and the gauge
invariant Wilson loop.  We give a detailed study of the Chern-Simons gauge theory on a
three-dimensional manifold.  In particular, we show how the WZW action is induced as a
boundary term under a gauge transformation of the Chern-Simons theory, and how the
wavefunctions in the Hilbert space of the Chern-Simons theory are recognized as the
conformal blocks that determine the $N$-point correlation functions of the WZW model.
In $\S$III we move on to knots and knot invariants.  We state the definitions that are
needed for the classification of topologically inequivalent closed curves embedded in
three-dimensional Euclidean space $\mathbb{R}^{3}$, and present the topological
manipulations that are permitted on the projections of knots onto a two-dimensional
plane (the Reidemeister moves).  We then define some of the relevant invariants - the
linking number, self-linking number (writhe), the Jones polynomial and the bracket
polynomial.  By defining the partition function for an arbitrary manifold as the
(unnormalized) expectation value of a Wilson loop with the exponential of the
Chern-Simons action as the measure, we are able to derive the skein relation for the
Jones polynomial, showing that the invariant has an intrinsic three-dimensional
definition.  We also derive the Gauss linking number which arises when the gauge
group is $U(1)$.  This completes the background that is neccessary to understand
the physical states of the quantum gravitational field.  In $\S$IV we describe the
second-order (metric), first-order (tetrad) and self-dual (connection) formulations
of general relativity with a non-zero cosmological constant.  We use the latter with
a densitized triad as conjugate momentum to derive the phase space which contains
seven primary constraints that vanish weakly on the constraint surface.  We define
the fundamental loop observables of the quantum theory and describe the spin networks
that arise in loop space.  We end the formal discussion with the Kodama state, in
particular how it is simultaneously a semiclassical state as well as an exact state,
how it is equivalent to the bracket polynomial of framed links in loop space and how
the $SU(2)$ spin networks are q-deformed due to the requirement of a framing of the
loop observables.

\section{Chern-Simons Theory and Conformal Field Theory}

\subsection{Loop States in Gauge Theory}

Let us begin by reminding the reader of the basic definitions of non-Abelian gauge
theories.  We have a vector potential (connection) one-form
$\bm{A}=A_{\mu}^{a}T^{a}dx^{\mu}$ that is valued in a Lie algebra with gauge group $G$
and generators $T^{a}$.  This connection determines the gauge covariant derivative
and curvature by
\bea
D_{\mu}\phi &=& \partial_{\mu}\phi + A_{\mu}^{a}T^{a}\phi\\
F_{\mu\nu} &=& [D_{\mu},D_{\nu}] 
           = \partial_{\mu}A_{\nu} - \partial_{\nu}A_{\mu} + [A_{\mu},A_{\nu}].
\label{curvature}
\eea
The curvature two-form $\bm{F}=(1/2)F_{\mu\nu}^{a}T^{a}dx^{\mu}\wedge dx^{\nu}$
is defined on an arbitrary four-dimensional manifold $\mathcal{M}$.  A gauge transformation
$g\in G$ is a zero-form whose action on the connection and curvature is given by
\bea
\tilde{\bm{A}} : \bm{A} = g^{-1}dg + g^{-1}\tilde{\bm{A}}g
\quad
\mbox{and}
\quad
\tilde{\bm{F}} : \bm{F} = g^{-1}\tilde{\bm{F}}g.
\label{trans}
\eea
The space of all connections is denoted $\mathcal{A}$, and the set of all gauge
transformations is said to form the group $\mathcal{G}$.  We write
$\mathcal{A}/\mathcal{G}$ for the space of gauge equivalent classes of connections,
i.e. the space of all connections modulo gauge transformations.  The action for pure
Yang-Mills theory \cite{hatfield,baemun} on $\mathcal{M}$ is
\bea
\mathcal{S}_{YM}[g_{\mu\nu},\bm{A}]
= \frac{1}{2}\int_{\mathcal{M}}\mbox{Tr}(\bm{F}\wedge\bm{F}^{*})
= -\frac{1}{4}\int_{\mathcal{M}}d^{4}x\sqrt{|\mbox{det}(g_{\mu\nu})|}
\mbox{Tr}(F_{\mu\nu}F^{\mu\nu}).
\label{ymaction}
\eea
Here $\mbox{Tr}$ is a non-degenerate invariant bilinear form on the Lie algebra
called the ``metric'' $\gamma^{ab}=p\mbox{Tr}(T^{a}T^{b})$ for some normalization
$p$ and $\bm{F}^{*}$ is the Hodge dual of $\bm{F}$.  Variation of the action with
respect to $\bm{A}$ yields the equations of motion $D\bm{F}^{*}=0$.

Now let $\gamma=\gamma(s)$ be a curve that is parametrized by $s$.  The path-ordered
product of the connection $\bm{A}$ with components $A_{\mu}=A_{\mu}(x(s))\in\mathcal{A}$
is defined such that
\bea
PA_{\mu}(x(s_{1}))\cdots A_{\mu}(x(s_{n}))=A_{\mu}(x(s_{1}))\cdots A_{\mu}(x(s_{n})),
\quad
s_{1}\geq\cdots\geq s_{n}.
\eea
This is a permutation of factors ordered so that larger values of $s_{i}$ appear left to right.
Now define on $\gamma$ the path-ordered exponential (called the Wilson line)
\bea
W[\gamma] = P\mbox{exp}\left[ig\int_{\gamma}dx^{\mu}A_{\mu}\right].
\eea
This is the holonomy of the connection $A_{\mu}$ on the principal
$G$-bundle.  It is required to transform as $W[\gamma] \rightarrow U(z)W[\gamma]U^{\dagger}(y)$
under a general gauge transformation $U$.  If we take $\gamma$ to be a closed loop then the
loop integral transforms as $W[\gamma]\rightarrow U(y)W[\gamma]U^{\dagger}(y)$, and taking the
trace of this gives
$\mbox{Tr}W[\gamma]\rightarrow\mbox{Tr}\left(U(y)W[\gamma]U^{\dagger}(y)\right)
=\mbox{Tr}W[\gamma]$.  This implies that the relevant quantity that is gauge invariant is
\bea
\mathcal{W}[\gamma] = \mbox{Tr}P\mbox{exp}\left[ig\oint_{\gamma}dx^{\mu}A_{\mu}\right],
\eea
called the Wilson loop \cite{hatfield}.  If $G=U(1)$ then the path ordering and trace are irrelevant
since the potentials commute.  In this case the Wilson loop gives a measure of the phase that is
acquired by a charged particle as it moves around the path $\gamma$ through an electromagnetic
field given by $F_{\mu\nu}=\partial_{\mu}A_{\nu}-\partial_{\nu}A_{\mu}$.

Let us now introduce some notation for the Wilson loops that we will be needing below.
If $\gamma_{i}$ represent segments of curves then $\gamma_{i}\circ\gamma_{j}$ says that
the beginning of $\gamma_{j}$ is glued onto the end of $\gamma_{i}$.  The segments
$\gamma_{i}$ and $\gamma_{j}$ may not be closed, but the composition of the two must form
a closed loop.  The segment $\lambda$ is necessarily an open curve.  Also the inverse
$\gamma^{-1}$ is the segment $\gamma$ with opposite orientation.  With this, the Wilson
loops satisfies the following properties \cite{loll,gampul}:
\bea
\mathcal{W}[0] &=& 1
\label{mandelstam1}\\
\mathcal{W}[\gamma] &=& \mathcal{W}[\gamma\circ\lambda\circ\lambda^{-1}]
\label{mandelstam2}\\
\mathcal{W}[\gamma_{1}\circ\gamma_{2}] &=& \mathcal{W}[\gamma_{2}\circ\gamma_{1}].
\label{mandelstam3}
\eea
In addition, the Wilson loops also satisfy two properties which hold only for gauge
group $G=SL(2,\mathbb{C})$.  These are
\bea
\mathcal{W}[\gamma] &=& \mathcal{W}[\gamma^{-1}]
\label{mandelstam4}\\
\mathcal{W}[\gamma_{1}]\mathcal{W}[\gamma_{2}]
&=& \frac{1}{2}\left(\mathcal{W}[\gamma_{1}\circ\lambda\circ\gamma_{2}\circ\lambda^{-1}]
    + \mathcal{W}[\gamma_{1}\circ\lambda\circ\gamma_{2}^{-1}\circ\lambda^{-1}]\right),
\label{mandelstam5}
\eea
The properties (\ref{mandelstam1})-(\ref{mandelstam5}) are called the Mandelstam
constraints.  Diagrammatically for the simple case where
$\gamma_{1}\sim{\raise6pt\xy 0;/r1pc/: \hcap-<=\hcap \endxy}$ and
$\gamma_{2}\sim{\raise6pt\xy 0;/r1pc/: \hcap-<=\hcap \endxy}$ (\ref{mandelstam5}) says that
\bea
\mathcal{W}[{\raise6pt\xy 0;/r1pc/: \hcap-<=\hcap \endxy}]
\mathcal{W}[{\raise6pt\xy 0;/r1pc/: \hcap-<=\hcap \endxy}]
= \frac{1}{2}
\left(\mathcal{W}[{\raise6pt\xy 0;/r1pc/: \hcap-<=\huntwist\hcap<= \endxy}]
+ \mathcal{W}[{\raise6pt\xy 0;/r1pc/: \hcap-<=\htwist\hcap>= \endxy}]\right).
\eea

\subsection{Chern-Simons and Wess-Zumino-Witten Actions}

Let $\bm{A}=A_{i}^{a}T^{a}dx^{i}$ be a connection one-form valued in a Lie algebra with gauge
group $G$ and generators $T^{a}$.  The Chern-Simons action on an arbitrary three-dimensional
manifold $\mathcal{M}_{3}$ is defined as
\bea
S_{CS}[\bm{A}] &=& \frac{k}{4\pi}\int_{\mathcal{M}_{3}}\mbox{Tr}
         \left(\bm{A}\wedge d\bm{A} + \frac{2}{3}\bm{A}\wedge\bm{A}\wedge\bm{A}\right)\nonumber\\
          &=& \frac{k}{8\pi}\int_{\mathcal{M}_{3}}\epsilon^{ijk}\mbox{Tr}
         \left(A_{i}(\partial_{j}A_{k}-\partial_{k}A_{j})+\frac{2}{3}A_{i}[A_{j},A_{k}]\right),
\label{chernsimons}
\eea
where $k$ is a coupling constant called the level.  The equations of motion obtained from
variation of the action with respect to $\bm{A}$ are $\bm{F}=d\bm{A}+\bm{A}\wedge\bm{A}=0$,
which means that the field strength $\bm{F}$ vanishes.  If $g\in G$ is a gauge transformation
defined by (\ref{trans}), the Chern-Simons action transforms as
\cite{jackiw,carlip1}
\bea
S_{CS}[\bm{A}] &=& S_{CS}[\tilde{\bm{A}}] - \frac{k}{4\pi}
            \int_{\partial\mathcal{M}_{3}}\mbox{Tr}\left[(dgg^{-1})\wedge\tilde{\bm{A}}\right]\nonumber\\
          &\phantom{=}&  - \frac{k}{12\pi}
            \int_{\mathcal{M}_{3}}\mbox{Tr}\left[(g^{-1}dg)\wedge(g^{-1}dg)\wedge(g^{-1}dg)\right].
\eea
If $\mathcal{M}_{3}$ is compact and without boundary then the boundary term above
vanishes, and $S_{CS}$ is gauge-invariant (modulo $2\pi$) if $k$ is an integer.
The last term is an integral multiple of $2\pi$ called the winding number of $g$.

If $\mathcal{M}_{3}$ has a boundary then a surface term needs to be added to the action to
ensure that $S_{CS}$ has proper extrema.  This is accomplished by imposing boundary
conditions on a field, i.e. a particular field is held fixed at $\partial\mathcal{M}_{3}$.
For two-dimensional manifolds there is a remarkable property that conformal structures
are the same as complex structures.  See, for example, Chapter 21 of Hatfield's book
\cite{hatfield}.  It is then natural to fix a complex structure on $\partial\mathcal{M}_{3}$
by  introducing a complex coordinate $z=x^{1}+ix^{2}$ with conjugate $\bar{z}=x^{1}-ix^{2}$
if the real coordinate system is given by $(x^{1},x^{2})$.  Choosing the axial gauge
$A_{0}=0$ results in a connection that is two-dimensional.  The components of this
connection transform as
\bea
A_{1} = A_{z}\pdiff{z}{x^{1}} + A_{\bar{z}}\pdiff{\bar{z}}{x^{1}} = A_{z} + A_{\bar{z}},
\quad
A_{2} = A_{z}\pdiff{z}{x^{2}} + A_{\bar{z}}\pdiff{\bar{z}}{x^{2}} = i(A_{z} - A_{\bar{z}})
\eea
so that in the holomorphic representation (in terms of $z$ and $\bar{z}$) the connection is written
as
$\bm{A}=A_{z}dz + A_{\bar{z}}d\bar{z}$.  Fixing the component $A_{z}$ on $\partial\mathcal{M}_{3}$
gives the action
\bea
S_{CS}^{\otimes}[\bm{A}] = \frac{k}{4\pi}\int_{\mathcal{M}_{3}}\mbox{Tr}
            \left(\bm{A}\wedge d\bm{A} + \frac{2}{3}\bm{A}\wedge\bm{A}\wedge\bm{A}\right)
            + \frac{k}{4\pi}\int_{\partial\mathcal{M}_{3}}dz\wedge d\bar{z}
            \mbox{Tr}(A_{z}A_{\bar{z}}).
\label{boundaryaction1}
\eea
Alternatively, fixing the component $A_{\bar{z}}$ on $\partial\mathcal{M}_{3}$ gives the action
\bea
S_{CS}^{\otimes}[\bm{A}] = \frac{k}{4\pi}\int_{\mathcal{M}_{3}}\mbox{Tr}
\left(\bm{A}\wedge d\bm{A} + \frac{2}{3}\bm{A}\wedge\bm{A}\wedge\bm{A}\right)
- \frac{k}{4\pi}\int_{\partial\mathcal{M}_{3}}dz\wedge d\bar{z}\mbox{Tr}(A_{z}A_{\bar{z}}).
\label{boundaryaction2}
\eea
When the action (\ref{boundaryaction1}) or the action (\ref{boundaryaction2}) are varied,
then the boundary terms cancel as can be verified.  Now, under the gauge transformation
$g\in G$ (defined by (\ref{trans})), the action (\ref{boundaryaction1}) transforms as
\bea
S_{CS}^{\otimes}[\bm{A}] &=& S_{CS}^{\otimes}[\tilde{\bm{A}}] + kS_{WZW}^{\oplus}[g,\tilde{A}_{z}]\\
S_{WZW}^{\oplus}[g,\tilde{A}_{z}] &=& \frac{1}{4\pi}\int_{\partial\mathcal{M}_{3}}\mbox{Tr}
(g^{-1}\partial_{z}gg^{-1}\partial_{\bar{z}}g - 2g^{-1}\partial_{\bar{z}}g\tilde{A}_{z})
+ \frac{1}{12\pi}\int_{\mathcal{M}_{3}}\mbox{Tr}(g^{-1}dg)^{3}.
\label{connection}
\eea
where $S_{WZW}^{\oplus}$ is the ``chiral'' Wess-Zumino-Witten action \cite{knizam,witten2} on
the boundary $\partial\mathcal{M}_{3}$ (see below for definition of chirality).  This establishes,
classically, the connection between Chern-Simons gauge theory in the bulk manifold and
conformal field theory on the boundary.

\subsection{Conformal Field Theory}

We will now discuss some general features of two-dimensional conformal field theory.  Our treatment here
is necessarily brief.  The main purpose is to introduce the conformal blocks which are essential for
the quantization of the Chern-Simons theory discussed above.  We follow the review article by
Gaberdiel \cite{gaberdiel}, to which the reader is referred for more details.

In $d$-dimensional Euclidean space $\mathbb{R}^{d}$, with $d\geq3$, the conformal group consists
of rotations and translations that preserve angles and lengths, the scale transformations given by
\bea
x^{\mu} \rightarrow \Omega x^{\mu},
\quad
x^{\mu} \in \mathbb{R}^{d},
\quad
\Omega \in \mathbb{R},
\eea
and the so-called special conformal transformations given by
\bea
x^{\mu} \rightarrow \frac{x^{\mu}+\bm{x}^{2}a^{\mu}}{1+2(\bm{x}\cdot\bm{a})+\bm{x}^{2}\bm{a}^{2}},
\quad
a^{\mu} \in \mathbb{R}^{d},
\quad
\bm{x}^{2} = x_{\mu}x^{\mu},
\quad
\bm{a}^{2} = a_{\mu}a^{\mu}.
\eea
For $d=2$, however, we can introduce the complex coordinates $z,\bar{z}$ for
$(x^{1},x^{2})\in\mathbb{R}^{2}$ such that $z=x^{1}+ix^{2}$ and $\bar{z}=x^{1}-ix^{2}$ as we did for
$\partial\mathcal{M}_{3}$ above.  Here a conformal transformation is defined simply by the analytic
map $z\rightarrow f(z)$ for some (locally) analytic function $f(z)$.  The conformal group in two
dimensions is therefore determined by the set of all analytic maps from the plane to itself, which
is infinite-dimensional.  A subgroup of such mappings form the set of M$\ddot{\mbox{o}}$bius
transformations given by
\bea
z \rightarrow f(z) = \frac{az+b}{cz+d},
\quad
a,b,c,d \in \mathbb{C},
\quad
ad-bc = 1.
\eea
To each such transformation can be associated a matrix
\bea
A =
\left(\begin{array}{cc}
        a & b
\\      c & d
\end{array}\right),
\quad
\mbox{det}(A) = 1.
\eea
This group is therefore isomorphic to the special linear group $SL(2,\mathbb{C})$.  In fact,
the M$\ddot{\mbox{o}}$bius group of automorphisms of the Riemann sphere is isomorphic to
$SL(2,\mathbb{C})/\mathbb{Z}_{2}$.  There is also an infinite set of infinitesimal
transformations given by
\bea
\map{\ell_{n}}{z}{z+\epsilon z^{n+1}},
\quad
\map{\bar{\ell}_{n}}{\bar{z}}{\bar{z}+\epsilon\bar{z}^{n+1}},
\quad
n\in\mathbb{Z}.
\eea
The generators $\ell_{n}=-z^{n+1}(d/dz)$ and $\bar{\ell}_{n}=-\bar{z}^{n+1}(d/d\bar{z})$
satisfy the commutation relations
\bea
[\ell_{m},\ell_{n}] = (m-n)\ell_{m+n}
\quad
\mbox{and}
\quad
[\bar{\ell}_{m},\bar{\ell}_{n}] = (m-n)\bar{\ell}_{m+n}.
\label{classcom}
\eea
These relations together with the commutator $[\ell_{m},\bar{\ell}_{n}]=0$ form an
infinite-dimensional Lie algebra called the Witt algebra or the classical Virasoro
algebra.  Upon quantization, the generators $\ell_{n},\bar{\ell}_{n}$ become operators
$L_{n},\bar{L}_{n}$, and the commutation relations (\ref{classcom}) suffer from ordering
ambiguities.  They require an extension by normal ordering, which occurs for the operators
$L_{0}$ and $\bar{L}_{0}$ (when $m=-n$).  The extended algebra is given by the commutation
relations
\bea
[L_{m},L_{n}] = (m-n)L_{m+n} + \frac{c}{12}m(m^{2}-1)\delta_{m,-n}
\eea
for the generators $L_{n}$, and
\bea
[\bar{L}_{m},\bar{L}_{n}] = (m-n)\bar{L}_{m+n} + \frac{c}{12}m(m^{2}-1)\delta_{m,-n}
\eea
for the generators $\bar{L}_{n}$.  These extended commutation relations together with
the $[L_{m},\bar{L}_{n}]=0$ form the quantum Virasoro algebra.  The constant $c$ is
called the central charge, or conformal anomaly.  It commutes with all the generators
$L_{n}$ and $\bar{L}_{n}$.

Like any quantum field theory, a two-dimensional conformal field theory is determined by the
space of states $\psi_{n}$ which form a Hilbert space $\mathcal{H}_{*}$, and the collection of
the correlation functions that are defined within some dense subspace
$\mathcal{F}\subset\mathcal{H}_{*}$.  The highest weight state of any conformal field theory is
the state $\psi$ whose $L_{0},\bar{L}_{0}$ eigenvalues are smallest; their eigenvalues
$L_{0}\psi=h\psi$ and $\bar{L}_{0}\psi=\bar{h}\psi$ are the conformal weights, and determine
the conformal transformation properties of $\psi$.  The sum $\Delta=h+\bar{h}$ is called the
scaling dimension, and the difference $s=h-\bar{h}$ is called the planar spin.  Let
$V(\psi;z,\bar{z})$ be a field associated to the state $\psi$ that satisfies
$\psi=V(\psi;0,0)\ket{0}$, and $\ket{0}$ is the ($SL(2,\mathbb{C})/\mathbb{Z}_{2}$-invariant)
``vacuum'' state.  Then under the conformal transformations $z\rightarrow f(z)$ and
$\bar{z}\rightarrow\bar{f}(\bar{z})$ the field $V(\psi;z,\bar{z})$ transforms as
\bea
V(\psi;z,\bar{z}) \rightarrow \left[\tdiff{f}{z}\right]^{h}\left[\tdiff{\bar{f}}{\bar{z}}\right]^{\bar{h}}
                              V(\psi;f(z),\bar{f}(\bar{z})).
\eea
A field that transforms in this way is called a primary field.  It turns out that there is a subspace
of states in $\mathcal{H}_{*}$ that transform as the vacuum state with respect to $L_{n}$ ($\bar{L}_{n}$).
These are known as chiral (anti-chiral) states.  The chiral (anti-chiral) fields then only depend on
$z$ ($\bar{z}$).  Therefore chiral fields are given by $V(\psi;z,\bar{z})=V(\psi;z)$ and anti-chiral
fields are given by $V(\psi;z,\bar{z})=V(\psi;\bar{z})$.  Correlation functions ($n$-point functions)
of the conformal field theory of $n$ primary fields $V(\psi_{n};z_{n},\bar{z}_{n})$ with conformal
dimensions $h_{n}$ and $\bar{h}_{n}$ are represented by
\bea
\bra{0}V(\psi_{1};z_{1},\bar{z}_{1})\cdots V(\psi_{n};z_{n},\bar{z}_{n})\ket{0}
= \langle V(\psi_{1};z_{1},\bar{z}_{1})\cdots V(\psi_{n};z_{n},\bar{z}_{n})\rangle.
\eea
The one-point function is given by
\bea
\langle V(\psi;z,\bar{z})\rangle = C
\eea
when $h=\bar{h}$ with $C$ a constant, and is zero otherwise; the two-point function is given by
\bea
\langle V(\psi_{1};z_{1},\bar{z}_{1})V(\psi_{2};z_{2},\bar{z}_{2})\rangle
= C(z_{1}-z_{2})^{-2h}(\bar{z}_{1}-\bar{z}_{2})^{-2\bar{h}}
\eea
when $h_{1}=\bar{h}_{1}$ and $h_{2}=\bar{h}_{2}$, and is zero otherwise.  Similar relations can be
found for the higher $n$-point functions.  In general the M$\ddot{\mbox{o}}$bius symmetry is used
to restrict these functions.  A remarkable feature of the WZW model is that the correlation
functions of chiral fields factorize holomorphically \cite{bpz}.  Let
$\Phi_{h_{\xi},\bar{h}_{\xi}}^{q_{\xi}\bar{q}_{\xi}}$ ($\xi\in\{1,2,\ldots,N\}$) be the primary
fields of the WZW model with conformal weights ($h_{\xi},\bar{h}_{\xi}$).  If the number $N$ is
finite then the conformal field theory is said to be rational.  In this case the fields are
tensor products of two finite-dimensional representations of a semi-simple Lie group with
left-hand and right-hand representations labelled by $q_{\xi}$ and $\bar{q}_{\xi}$,
respectively.  An $N$-point correlation function of the primary fields is then
\bea
\left\langle\Phi_{h_{1},\bar{h}_{1}}^{q_{1}\bar{q}_{1}}(\psi;z_{1},\bar{z}_{1})
\cdots\Phi_{h_{N},\bar{h}_{N}}^{q_{N}\bar{q}_{N}}(\psi;z_{N},\bar{z}_{N})\right\rangle
= \sum_{a,b}C^{ab}\mathcal{F}_{a}^{q_{1}\cdots q_{N}}(z_{1},\ldots,z_{N})
\bar{\mathcal{F}}_{b}^{\bar{q}_{1}\cdots\bar{q}_{N}}(\bar{z}_{1},\ldots,\bar{z}_{N})\nonumber\\
\eea
with constants $C^{ab}$ that are related to the structure constants of the operator algebra.
The functions $\mathcal{F}(z_{1},\ldots,z_{N})$ are the conformal blocks and are completely
determined by the conformal symmetry.

\subsection{Canonical Quantization of Chern-Simons Theory}

Now, we have seen how the Chern-Simons gauge theory induces the WZW theory on
$\partial\mathcal{M}_{3}$.  The relationship is far more profound when this theory
is quantized.  We now proceed to demonstrate this relationship.  To this end, let us
fix the topology of the three-manifold to $\mathcal{M}_{3}=\sigma\times R^{1}$, and
let $A_{0}=0$.  Here $A_{0}$ is the component of the connection $\bm{A}$ in the $R^{1}$-direction.
In this gauge the Chern-Simons action (\ref{chernsimons}) becomes
\bea
S_{CS}[\bm{A}] = \frac{k}{8\pi}\int dt\int_{\sigma}\epsilon^{ij}\mbox{Tr}\left(A_{i}
                 \frac{d}{dt}A_{j}\right),
\label{csgauged}
\eea
and there is one primary constraint $\fsent{S_{CS}}{A_{0}}=\epsilon^{ij}F_{ij}=0$.  The next step
would be to do a Legendre transform to obtain a function that is first order in time derivatives.
However, the Lagrangian above is already first order in time derivatives and the connections
$A_{i}$ and $A_{j}$ are thus canonically conjugate to each other.  Now recall from the definition
of the Poisson bracket (see Appendix B) that the canonical coordinates
$\{q^{\alpha},p_{\alpha}\}_{\alpha=1}^{n}$ for a discrete classical system with a finite number
of degrees of freedom will satisfy $\poisson{q^{\alpha}}{p_{\alpha}}=1$.  For a continuous system
with phase space defined by fields $\{\phi^{\alpha},\pi_{\alpha}\}_{\alpha=1}^{n}$, this
generalizes to $\poisson{\phi^{\alpha}(x)}{\pi_{\alpha}(y)}=\delta^{d}(x-y)$ (in $d$ dimensions).
For the gauge-fixed  action (\ref{csgauged}) we can read off the (equal-time) Poisson brackets as
\bea
\poisson{A_{i}^{a}(x)}{A_{j}^{b}(y)} = \frac{4\pi}{k}\epsilon_{ij}\delta^{ab}\delta^{2}(x-y).
\eea
Now we impose the constraints $F_{ij}=0$.  This leads to a finite-dimensional phase space given
by the configuration space of gauge equivalent classes of flat connections.  In other words, if
$\Gamma$ is the infinite-dimensional phase space of $\mathcal{A}/\mathcal{G}$, then the constraint
surface $\Gamma_{0}\subset\Gamma$ is the configuration space of $\mathcal{A}_{0}/\mathcal{G}$.
Witten calls this configuration space ``a flat bundle on moduli space''.  We can assume a complex
structure on $\sigma$ which then amounts to the separation of the connections $A_{i}$ into
holomorphic and anti-holomorphic degrees of freedom, as was done above for (\ref{boundaryaction1}).
The configuration space $\mathcal{A}_{0}/\mathcal{G}$ is then a flat vector bundle on moduli space.
It turns out that these bundles are exactly those that arise naturally in the holomorphic
factorization of the $N$-point functions that we have already defined above as the conformal
blocks $\mathcal{F}(z_{1},\ldots,z_{N})$.  We now see the deeper relationship between a generally
covariant gauge theory in three dimensions and a conformally invariant quantum field theory in two
dimensions: quantization of the Chern-Simons action on $\mathcal{M}_{3}=\sigma\times R^{1}$ produces
a Hilbert space $\mathcal{H}_{0}$ of wavefunctions that are recognized as the generating functionals
for two-dimensional current correlator blocks.

\section{The Theory of Knots}

\subsection{Topological Preliminaries}

A knot is a submanifold of $\mathbb{R}^{3}$ that is diffeomorphic to the circle $S^{1}$.
The simplest example of a knot is the unknot, a circle that contains zero crossings:
\bea
\{(x,y,z) \in \mathbb{R}^{3} | x^{2} + y^{2} = 1 , z = 0 \}
\quad
\sim
\quad
~\xygraph{
!{0;/r1pc/:}
!{\vcap-}
!{\vcap}
}~.
\eea
Another example of a knot is the trefoil knot
$\xygraph{
!{0;/r0.5pc/:}
!P3"a"{~>{}}
!P9"b"{~:{(1.3288,0):}~>{}}
!P3"c"{~:{(2.5,0):}~>{}}
!{\vover~{"b2"}{"b1"}{"a1"}{"a3"}}
!{"b4";"b2"**\crv{"c1"}}
!{\vover~{"b5"}{"b4"}{"a2"}{"a1"}}
!{"b7";"b5"**\crv{"c2"}}
!{\vover~{"b8"}{"b7"}{"a3"}{"a2"}}
!{"b1";"b8"**\crv{"c3"}}
"b0"[ddrr] 
}~$.
This is an example of an alternating knot where the crossings alternate between over
and under.  A link is a submanifold of $\mathbb{R}^{3}$ that is diffeomorphic to a
disjoint union of circles.  The circles are the components of the link.  A knot is
a link with one component.  An example of a link is the Hopf link
$\xygraph{
[uuuu]!{0;/r0.5pc/:}
[u]!{\vover}
!{\vover-}
[uur]!{\hcap[2]}
[l]!{\hcap[-2]}
}~$.
Two knots $\bm{K}$ and $\bm{K}^{\prime}$ are said to be ambient isotopic iff $\bm{K}$
can be continuously deformed into $\bm{K}^{\prime}$ through embeddings of $\mathbb{R}^{3}$.
An equivalence class of links with ambient isotopy as the equivalence relation is said to be
an isotopy class.  A knot is amphicheiral iff it is isotopic to its mirror image.  The
trefoil, for example, is not amphicheiral.  Also, a knot decomposes if it is the
connected sum of two knots.  A knot $\bm{K}$ is said to be the connected sum of knots
$\bm{K}_{1}$ and $\bm{K}_{2}$, denoted $\bm{K}_{1}\#\bm{K}_{2}$, if a two-sphere can be
embedded in $\mathbb{R}^{3}$ so that the two-sphere intersects $\bm{K}$ in exactly two
points so that $\bm{K}$ is bisected into pieces one of which is $\bm{K}_{1}$ minus a
small arc and $\bm{K}_{2}$ minus a small arc.  A knot is then prime if it cannot be
decomposed into two simpler knots.

It turns out that two knots $\bm{K}_{1}$ and $\bm{K}_{2}$ are isotopic iff there exists a
sequence of elemetary isotopies (called Reidemeister moves) between projection diagrams of
$\bm{K}_{1}$ and $\bm{K}_{2}$.  The Reidemeister moves are given in diagrammatic representation
in Fig. 1.  The first move RI consists of modifying a diagram by putting a twist in the
neighbourhood of a single strand.  The second move RII consists of cancelling a pair of crossings.
The third move consists of sliding a strand over a crossing (RIIIa) or under a crossing (RIIIb).
There is also a ``zeroth'' move R0 which consists simply of an isotopy of the plane.  We note that
these moves are local; they are applied (one at a time) to one segment of a knot diagram while
leaving the rest of the knot diagram unchanged.  The equivalence relation generated by all
Reidemeister moves except RI is called regular isotopy.  Also, a loop that cannot be deformed into
the unknot with the Reidemeister moves is said to be knotted.
\begin{figure}[t]
\centering
\includegraphics[width=1.5in]{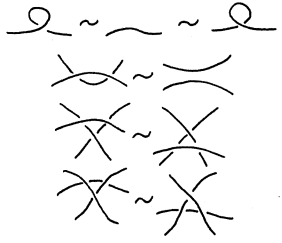}
\caption{The Reidemeister moves RI-RIII(a,b).  Here a strand is a single line segment.}
\end{figure}

\subsection{Link Invariants}

By Reidemeister's theorem \cite{reidemeister}, the topological properties of knots
and links in three-space ($\mathbb{R}^{3}$ or $S^{3}$) can be studied by working with
diagrams of the knots taken up to the Reidemeister moves.  That is, if diagrams for two
knots or links are ambient isotopic in three-dimensional space iff the diagrams can be
obtained one from another by a sequence of Reidemeister moves.

A knot or link is said to be oriented if a direction along the curve is chosen for each
component of the knot or link.  When we have a diagram of an oriented knot or link, then
each crossing inherits and orientation sign.  We define for right-handed orientation
positive sign $\epsilon(~{\raise6pt\xy 0;/r1pc/: \xoverh=> \endxy}~) = +1$ and for
left-handed orientation negative sign
$\epsilon(~{\raise6pt\xy 0;/r1pc/: \xunderh=> \endxy}~) = -1$.  Let us first consider
the self-linking number.  For an oriented (one-component) knot diagram $\bm{K}$ with
$N$ crossings $\{1,2,\ldots,N\}$ this is
\bea
w(\bm{K}) &=& \sum_{q=1}^{N}\epsilon_{q}.
\eea
We note that this function is not preserved under RI but is preserved by the
other Reidemeister moves, and is therefore a regular isotopy invariant.  Now let us consider
an oriented diagram formed by two knots $\bm{K}_{1}$ and $\bm{K}_{2}$.  The linking number
is
\bea
L(\bm{K}_{1},\bm{K}_{2}) &=& \frac{1}{2}\sum_{p\in\bm{K}_{1}\cap\bm{K}_{2}}\epsilon(p),
\eea
where the intersection $\bm{K}_{1}\cap\bm{K}_{2}$ does not include the self-crossings of
$\bm{K}_{1}$ or $\bm{K}_{2}$.  As an example we will calculate the linking number for Hopf
links with opposite orientations.  For
$\bm{K}_{1}\sim ~{\raise6pt\xy 0;/r1pc/: \hcap->=\hcap \endxy}$ and
$\bm{K}_{2}\sim ~{\raise6pt\xy 0;/r1pc/: \hcap->=\hcap \endxy}$ so that
$\bm{K}_{1}\cup\bm{K}_{2}\sim\xygraph{[uuuu]!{0;/r0.6pc/:}[u]!{\vover}!{\vover-}
[uur]!{\hcap[2]@(0)=>}[l]!{\hcap[-2]@(0)=>}}$ we have
\bea
L(\bm{K}_{1},\bm{K}_{2}) = \frac{1}{2}\left(
\epsilon(\; {\raise6pt\xy 0;/r1pc/: \xoverh=> \endxy}\; )
+ \epsilon(\; {\raise6pt\xy 0;/r1pc/: \xoverh=> \endxy}\; )\right) = +1,
\eea
while for $\bm{K}_{1}\sim~{\raise6pt\xy 0;/r1pc/: \hcap-<=\hcap \endxy}$ and
          $\bm{K}_{2}\sim~{\raise6pt\xy 0;/r1pc/: \hcap->=\hcap \endxy}$ so that
$\bm{K}_{1}\cup\bm{K}_{2}\sim\xygraph{[uuuu]!{0;/r0.6pc/:}[u]!{\vover}!{\vover-}
[uur]!{\hcap[2]@(0)=>}[l]!{\hcap[-2]@(0)=<}}$ we have
\bea
L(\bm{K}_{1},\bm{K}_{2}) = \frac{1}{2}\left(
\epsilon(\; {\raise6pt\xy 0;/r1pc/: \xunderh=> \endxy}\; )
+ \epsilon(\; {\raise6pt\xy 0;/r1pc/: \xunderh=> \endxy}\; )\right) = -1.
\eea

Before we go on to define the bracket polynomial we will introduce the notion of framing.
Let $\bm{K}$ be a knot that is defined by a closed curve $C$.  The framing of $\bm{K}$
is a vector field that is normal to the curve $C$, whose endpoints generate a boundary
curve $C^{\prime}$.  So a framed knot can be visualized as a closed ribbon with boundaries
defined by the curves $C$ and $C^{\prime}$.  The first Reidemeister move RI fails for
framed knots. Instead, a curl in a diagram corresponds to a $2\pi$ twist in the ribbon.
See Fig. 2.  We find that both $[\mbox{a}]$ and $[\mbox{b}]$ represent the same $2\pi$
twist and so we set $[\mbox{a}]\sim[\mbox{b}]$.
\begin{figure}[t]
\centering
\includegraphics[width=1.5in]{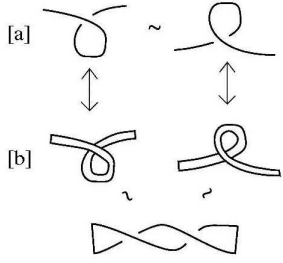}
\caption{The modified reidemeister move RI$^{\prime}$ for a framed knot segment.}
\end{figure}
This modified deformation is henceforth labelled RI$^{\prime}$.

Now we are ready to define the bracket polynomial.  Let $\bm{\mathcal{L}}$ be a
multi-component link containing $N$ crossings.  A state of $\bm{\mathcal{L}}$ is
obtained by choosing a smoothing of each crossing.  Assigned to each crossing of
$\bm{\mathcal{L}}$ is a choice $\sigma_{q}\in\{A,B\}$, where the action of the
vertex weights $A,B$ on the $q$th crossing is given by
\bea
{\raise6pt\xy 0;/r1pc/: \xoverh \endxy}~
\rightarrow
~A(~{\raise6pt\xy 0;/r1pc/: \xunoverh \endxy}~)
\quad
\mbox{and}
\quad
{\raise6pt\xy 0;/r1pc/: \xoverh \endxy}~
\rightarrow
~B(~{\raise6pt\xy 0;/r1pc/: \xunoverv \endxy}~);
\eea
smoothing the crossing horizontally results in a multiplicative factor of $A$ while smoothing
the crossing vertically results in a multiplicative factor of $B$.  The resulting state will
be a finite set of disjoint circles of cardinality $\bar{\sigma}$.  The bracket polynomial is
then a state sum given by
\bea
\langle~\bm{\mathcal{L}}~\rangle = \sum_{\sigma}d^{\bar{\sigma}-1}\prod_{q}\sigma_{q}.
\label{kauffman1}
\eea
This is a polynomial in the commuting variables $A$, $B$ and $d$.  The skein relations
that define the bracket polynomial are
\bea
\langle ~{\raise6pt\xy 0;/r1pc/: \xoverh \endxy}~ \rangle
= A \langle ~{\raise6pt\xy 0;/r1pc/: \xunoverh \endxy}~ \rangle +
  B \langle ~{\raise6pt\xy 0;/r1pc/: \xunoverv \endxy}~ \rangle
\quad
\mbox{and}
\quad
\langle
~\xygraph{
!{0;/r1pc/:}
!{\vcap-}
!{\vcap}
}~ \cup~ \bm{K}~ \rangle
= d\langle ~\bm{K}~ \rangle.
\label{kauffman2}
\eea
These equations follow directly from the definition of the state sum, where a normalization of
unity has been employed for the unknot.  Here $\bm{K}$ does not intersect the added loop.  Note
that the bracket polynomial is an invariant of regular isotopy; if the axioms are applied to a
twist then
\bea
\langle
~\xygraph{
[uuuu]!{0;/r0.6pc/:}
[d(0.6)]!{\xcaph[0.3]@(0)}
[u(1)l(0.7)]!{\vunder-}
[r(1)]!{\xcaph[0.3]@(0)}
[ull]!{\vcap}
}~
\rangle = (Bd+A) \langle
~\xygraph{
!{0;/r1pc/:}
!{\xcaph[1]@(0)}
}~
\rangle
\quad
\mbox{and}
\quad
\langle
~\xygraph{
[uuuu]!{0;/r0.6pc/:}
[d(0.6)]!{\xcaph[0.3]@(0)}
[u(1)l(0.7)]!{\vover-}
[r(1)]!{\xcaph[0.3]@(0)}
[ull]!{\vcap}
}~
\rangle = (Ad+B) \langle
~\xygraph{
!{0;/r1pc/:}
!{\xcaph[1]@(0)}
}~
\rangle.
\label{break}
\eea
In fact, the bracket polynomial is invariant under the framed Reidemeister moves for
a specific choice of the commuting variables $A,B,d$.  As defined, the bracket polynomial
is invariant under RI$^{\prime}$, but fails for RII:
\bea
\langle
~\xygraph{
!{0;/r0.5pc/:}
[u(0.8)]
!{\xcaph@(0)}
!{\vunder}
!{\vover-}
[l]!{\xcaph@(0)}
[r]!{\xcaph@(0)}
[uul]!{\xcaph@(0)}
}~
\rangle
= AB\langle ~{\raise6pt\xy 0;/r1pc/: \xunoverv \endxy}~ \rangle
  +\left[A^{2}+ABd+B^{2}\right]\langle ~{\raise6pt\xy 0;/r1pc/: \xunoverh \endxy}~ \rangle.
\eea
Invariance requires that $AB=1$ and $A^{2}+ABd+B^{2}=0$ which uniquely determines the
variables such that $B=A^{-1}$ and $d=-(A^{2}+A^{-2})$.  With this choice the bracket
polynomial is now also invariant under the action of RIII.  Now we can write the defining
state sum (\ref{kauffman1}) for framed links that is invariant under the Reidemeister moves
as a polynomial in the single variable $A$, with the skein relations
\bea
\langle ~{\raise6pt\xy 0;/r1pc/: \xoverh \endxy}~ \rangle
= A \langle ~{\raise6pt\xy 0;/r1pc/: \xunoverh \endxy}~ \rangle +
  A^{-1} \langle ~{\raise6pt\xy 0;/r1pc/: \xunoverv \endxy}~ \rangle
\quad
\mbox{and}
\quad
\langle 
~\xygraph{
!{0;/r1pc/:}
!{\vcap-}
!{\vcap}
}~ \cup~ \bm{K}~ \rangle
= -(A^{2}+A^{-2})\langle ~\bm{K}~ \rangle.
\label{kauffman3}
\eea
Let us now define the Jones polynomial.  For a framed oriented link $\bm{\mathcal{L}}$
this is a polynomial in $A$ such that
\bea
V_{\bm{\mathcal{L}}}(A) = (-A^{-3})^{w(\bm{\mathcal{L}})}\langle ~\bm{\mathcal{L}}~ \rangle (A).
\eea
The skein relations for this polynomial (with $q=A^{-4}$) are
\bea
q ~{\raise6pt\xy 0;/r1pc/: \xoverh=> \endxy}
- q^{-1} ~{\raise6pt\xy 0;/r1pc/: \xunderh=> \endxy}
= (q^{\frac{1}{2}}-q^{-\frac{1}{2}}) ~{\raise6pt\xy 0;/r1pc/: \xunoverh=> \endxy}~
\quad
\mbox{and}
\quad
{\raise6pt\xy 0;/r1pc/: \hcap-<=\hcap \endxy} = 1.
\eea
Now note the relation for the bracket polynomial under the action of RI for unframed
links given by (\ref{break}), with coefficients $Bd+A=-A^{-3}$ and $Ad+B=-A^{3}$.  Under RI
this extra factor acquired by the bracket polynomial is cancelled by the coefficient in the
definition of the Jones polynomial above.  The Jones polynomial is therefore an invariant of
oriented links.  The framing is irrelevant.

\subsection{Link Invariants from Gauge Theory}

The Jones polynomial was originally defined for knots and links in the three-sphere $S^{3}$.
We will proceed to show that the exponential of the Chern-Simons action as a measure provides an
intrinsic definition of the Jones polynomial in an arbitrary three-dimensional manifold
$\mathcal{M}_{3}$.  The reader should note that the functional integral so defined is only
understood heuristically.  It is nevertheless, a powerful heuristic device, leading to a host
of results.  We might expect this because the Chern-Simons action is invariant under diffeomorphisms,
so a change of basis to loop space (see below, where the loop transform is defined) should give a knot
invariant for every gauge group $G$.  Let $C$ be an oriented closed curve in $\mathcal{M}_{3}$.  We
consider the Wilson loop
\bea
\mathcal{W}_{R}[C] = \mbox{Tr}_{R}P\mbox{exp}\left[\oint_{C}A_{i}dx^{i}\right]
\eea
obtained by calculating the holonomy of the connection $A_{i}$ in the representation $R$ of the gauge
group $G$.  Clearly the closed curve $C$ can be a knot; taking $r$ oriented and non-intersecting knots
$C_{\xi}$ ($\xi\in\{1,2,\ldots,r\}$) whose union is a link $\bm{\mathcal{L}}$, we define the partition
function of the manifold $\mathcal{M}_{3}$ such that
\bea
\mathcal{Z}(\bm{\mathcal{L}}) = \int_{\mathcal{A}/\mathcal{G}}(DA)
              \mbox{exp}(iS_{CS})\prod_{\xi=1}^{r}\mathcal{W}_{R_{\xi}}(C_{\xi}).
\label{partition}
\eea
This is a path integral over all gauge orbits where we have assigned a representation $R_{\xi}$
of $G$ to each knot $C_{\xi}$.  The partition function above is the unnormalized expectation value
of the given Wilson loop.  Evaluating this path integral amounts to the determination of a vector
$\psi$ in the physical Hilbert space $\mathcal{H}_{\mathcal{M}_{3}}$, i.e. the space of
gauge-invariant wavefunctions in the presence of Wilson loops.  We will now consider the special
case $\mathcal{M}_{3}=S^{3}$ and $G=SU(N)$.  Assume that there is a link present in $S^{3}$ and we
cut $S^{3}$ into two sections $\mathcal{M}_{L}$ and $\mathcal{M}_{R}$.  The boundaries of these
sections will now be Riemann surfaces intersecting the link transversely in a set of marked points.
In the case where there are four marked points and the Wilson lines are all in the same
$N$-dimensional representation of $SU(N)$, then (we state without proof that) the physical Hilbert
space will be two-dimensional.  If the sections $\mathcal{M}_{L}$ and $\mathcal{M}_{R}$ are pasted
back together, then the resulting link invariant may be the original, or may be a result of pasting
$\mathcal{M}_{L}$ with $\mathcal{X}_{1}$ or with $\mathcal{X}_{2}$.  This is illustrated in Fig. 3.
\begin{figure}[t]
\centering
\includegraphics[width=2.8in]{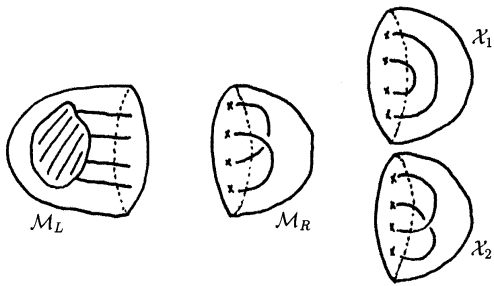}
\caption{Cutting of the manifold $S^{3}$ with Wilson loops into two manifolds $\mathcal{M}_{L}$
and $\mathcal{M}_{R}$.}
\end{figure}

Now let $\psi_{L}\in\mathcal{H}_{\mathcal{M}_{L}}$ and
$\psi_{R},\psi_{1},\psi_{2}\in\mathcal{H}_{\mathcal{M}_{R}}$.  Here $\psi_{R}$ is a vector associated
to the boundary of $\mathcal{M}_{R}$, $\psi_{1}$ is a vector associated to the boundary of
$\mathcal{X}_{1}$ and $\psi_{2}$ is a vector associated to the boundary of $\mathcal{X}_{2}$.
We note that any three vectors in a two-dimensional vector space are linearly dependent, and since
$\psi_{R},\psi_{1},\psi_{2}\in\mathcal{H}_{\mathcal{M}_{R}}$ are vectors contained in the same
Hilbert space we have that
\bea
\alpha\psi_{R} + \beta\psi_{1} + \gamma\psi_{2} = 0.
\label{dependence}
\eea
Here the constants $\alpha,\beta,\gamma\in\mathbb{C}^{1}$.
We may then postulate a natural pairing of vectors $(\psi_{L},\psi_{R})$ to the partition function
$\mathcal{Z}(\bm{\mathcal{L}})$, $(\psi_{L},\psi_{1})$ to the partition function
$\mathcal{Z}(\bm{\mathcal{L}}_{1})$ and $(\psi_{L},\psi_{2})$ to the partition function
$\mathcal{Z}(\bm{\mathcal{L}}_{2})$.  The relation (\ref{dependence}) becomes
\bea
\alpha(\psi_{L},\psi_{R}) + \beta(\psi_{L},\psi_{1}) + \gamma(\psi_{L},\psi_{2}) = 0,
\eea
and this translates into a reccursion relation for the expectation values of the Wilson loops given by
\bea
\alpha\mathcal{Z}(\bm{\mathcal{L}}) + \beta\mathcal{Z}(\bm{\mathcal{L}}_{1}) 
+ \gamma\mathcal{Z}(\bm{\mathcal{L}}_{2}) = 0.
\label{translation1}
\eea
Diagrammatically this is interpreted as
\bea
\alpha ~{\raise6pt\xy 0;/r1pc/: \xoverh=> \endxy}
+ \beta ~{\raise6pt\xy 0;/r1pc/: \xunoverh=> \endxy}
+ \gamma ~{\raise6pt\xy 0;/r1pc/: \xunderh=> \endxy} = 0.
\label{translation2}
\eea
Now, the explicit evaluation of the constants $\alpha,\beta,\gamma$ would divert us too much
from the main purpose of this paper, but it is neccessary to state the final result.  The reader
is referred to Witten's paper \cite{witten1} and references therein for details of the evaluation.
For a connection $\bm{A}$ of Chern-Simons theory with gauge group $G=SU(N)$ and all Wilson lines
in the same $N$-dimensional representation of $SU(N)$, we have that
\bea
\alpha &=& -\mbox{exp}\left(\frac{2\pi i}{N(N+k)}\right)\\
\beta &=& -\mbox{exp}\left(\frac{i\pi(2-N-N^{2})}{N(N+k)}\right)
        + \mbox{exp}\left(\frac{i\pi(2+N-N^{2})}{N(N+k)}\right)\\
\gamma &=& \mbox{exp}\left(\frac{2\pi i(1-N^{2})}{N(N+k)}\right).
\eea
Multiply these by a factor $\mbox{exp}[\pi i(N^{2}-2)/N(N+k)]$ and define a variable
$q\equiv\mbox{exp}[2\pi i/(N+k)]$.  Now (\ref{translation1}) and (\ref{translation2}) become
\bea
-q^{\frac{N}{2}}\mathcal{Z}(\bm{\mathcal{L}}) 
    + (q^{\frac{1}{2}}-q^{-\frac{1}{2}})\mathcal{Z}(\bm{\mathcal{L}}_{1})
+ q^{-\frac{N}{2}}\mathcal{Z}(\bm{\mathcal{L}}_{2}) &=& 0\\
-q^{\frac{N}{2}} ~{\raise6pt\xy 0;/r1pc/: \xoverh=> \endxy}
+ (q^{\frac{1}{2}}-q^{-\frac{1}{2}}) ~{\raise6pt\xy 0;/r1pc/: \xunoverh=> \endxy}
+ q^{-\frac{N}{2}} ~{\raise6pt\xy 0;/r1pc/: \xunderh=> \endxy} &=& 0.
\eea
This is the skein relation for the HOMFLY polynomial \cite{fyhlmo}, which is the
generalization of the Jones polynomial to $G=SU(N)$.  If $G=SU(2)$ we get the skein
relation for the Jones polynomial.

Before moving on to gravitation, we will briefly discuss a simple case of the partition
function (\ref{partition}) in order to illustrate a subtlety that arises when evaluating the
path integral (\ref{partition}) for a specific case.  Take $G=U(1)$ and $\mathcal{M}_{3}=S^{3}$.
The potentials commute so the cubic term $\bm{A}\wedge\bm{A}\wedge\bm{A}$ in the Chern-Simons
action vanishes, while the trace and path-ordering become irrelevant.  Consider in $S^{3}$ two
non-intersecting knots $C_{a}$ and $C_{b}$.  To the two different representations of $U(1)$ we
assign integers $n_{a}$ and $n_{b}$.  Evaluating the integral along the knots $C_{a}$ and $C_{b}$
in a region of $S^{3}$ that is essentially Euclidean space with $x^{i}$ and $y^{j}$ the coordinates
along the knots, the partition function for the case $a\neq b$ becomes
\bea
\mathcal{Z}(C_{a},C_{b}) = \frac{1}{4k}\int_{C_{a}}dx^{i}\int_{C_{b}}dy^{j}
                           \epsilon_{ijk}\frac{x^{k}-y^{k}}{|x-y|^{3}}.
\eea
This is the Gauss linking number, which was first written down by Gauss more than one hundred
and fifty years ago!  It is well defined provided that the two knots do not touch or intersect,
because the integral diverges at the point $x=y$.  Witten regularized this divergence for the
case $a=b$ by giving the knots a framing, and interpreted the resulting quantity as the
self-linking number of one knot.  Actually, this framing regularization is neccessary for all
gauge groups.  We have not mentioned it before because it was not relevant to the derivation
of the skein relation, i.e. we did not need to evaluate any integrals.  In a sense, we found
a reccursion relation for Wilson loop expectations for unframed loops, thereby neglecting the
self-linking.  That is why we obtained the Jones Polynomial.  Later we will see that the need
for framing will arise naturally for a state of quantum gravity with non-zero cosmological
constant.  There we will obtain the bracket polynomial for $G=SU(2)$, which depends on the
framing.

\section{The Gravitational Field: Classical Theory}

\subsection{Second Order Formulation}

The use of tensors in general relativity is motivated by the principle of general covariance
\cite{wald,carroll}, which states that the laws of physics must be independent of the
coordinates being used.  For example, the metric tensor $g_{\mu\nu}$ relates distances in space
and time via the invariant line interval $ds^{2}=g_{\mu\nu}dx^{\mu}dx^{\nu}$, and may be seen
as a generalization of the Pythagorean theorem to curved four-dimensional spacetime.  The
metric tensor is normally taken to be the dynamical field of the gravitational force.  The
equations relating the gravitational field to its source are the Einstein equations
\bea
R_{\mu\nu} - \frac{1}{2}Rg_{\mu\nu} - \Lambda g_{\mu\nu} = 8\pi G_{N}T_{\mu\nu}.
\label{geinstein}
\eea
Here the Ricci tensor $R_{\mu\nu}$ encodes information about the curvature, and its contraction
$R=g^{\mu\nu}R_{\mu\nu}$ is the Ricci scalar.  The material contents of the system, whatever it
might be, are described by the stress-energy tensor $T_{\mu\nu}$.  The quantity $\Lambda$ is
the cosmological constant, which manifests itself as an acceleration whose magnitude increases
with distance $r$ as $|\Lambda|r/3$ in the weak-field limit.  The dimensions carried by $\Lambda$
are $L^{-2}$, which implies that this term must be significant globally.  The Einstein equations
(\ref{geinstein}) can be derived from an action, either the Einstein-Hilbert action
\bea
S[g_{\mu\nu}] = \frac{1}{16\pi G_{N}}\int d^{4}x\sqrt{|\mbox{det}(g_{\mu\nu})|}
                (R[g_{\mu\nu}]-2\Lambda) + S_{matter}
\label{ehaction}
\eea
by taking the Ricci scalar to be a functional of the metric, or equivalently the Hilbert-Palatini
action
\bea
S[g_{\mu\nu},\Gamma_{\phantom{a}\mu\nu}^{\gamma}]
= \frac{1}{16\pi G_{N}}\int d^{4}x\sqrt{|\mbox{det}(g_{\mu\nu})|}
  (R[\Gamma_{\phantom{a}\mu\nu}^{\gamma}]-2\Lambda) + S_{matter}
\label{hpaction}
\eea
by taking the Ricci scalar to be a functional of the Christoffel coefficients of the second kind
$\Gamma_{\phantom{a}\mu\nu}^{\gamma}$.  The matter action
$S_{matter}=S_{matter}[g_{\mu\nu},\phi,\psi,\bm{A}]$ is generally a function of the scalar field
$\phi$, spinor field $\psi$ and vector field $\bm{A}$ in addition to the metric $g_{\mu\nu}$.
The stress-energy tensor is obtained by varying the matter action with respect to the metric
such that
\bea
T_{\mu\nu} = -\frac{2}{\sqrt{-g}}\fdiff{S_{matter}}{g^{\mu\nu}}.
\eea
It must be noted, however, that this tensor is not the same as the canonical stress-energy
tensor that is obtained from the symmetries of the Lagrangian, i.e. Noether's theorem.
The equivalence only holds for a scalar field (up to a numerical factor).  In general,
the canonical stress-energy tensor will not even be symmetric, and in many cases such as
electrodynamics, will not even be gauge invariant.  Thus we define the stress-energy tensor
as that which is obtained by variation of the matter action with respect to the gravitational
field.  When the action (\ref{ehaction}) and the action (\ref{hpaction}) are varied with
respect to $g_{\mu\nu}$ we get the Einstein equations.  The virtue of using (\ref{hpaction})
over (\ref{ehaction}), however, is that variation of the former with respect to
$\Gamma_{\phantom{a}\mu\nu}^{\gamma}$ also gives the metric compatability condition
$\nabla_{\gamma}g_{\mu\nu}=0$.

\subsection{First Order Formulation}

The Einstein theory can also be formulated in terms of an orthonormal tetrad frame
$e_{\mu}^{\phantom{a}I}$ that satisfies
\bea
g^{\mu\nu}e_{\mu}^{\phantom{a}I}e_{\nu}^{\phantom{a}J} = \eta^{IJ}
\quad
\mbox{and}
\quad
\eta_{IJ}e_{\mu}^{\phantom{a}I}e_{\nu}^{\phantom{a}J} = g_{\mu\nu}.
\eea
Locally then the manifold is flat Minkowski space $\mathbb{M}^{4}$ and the curvature
is encoded in the tetrad frame field.  The covariant derivative acts on spacetime indices in the
usual way:
\bea
\nabla_{\mu}v^{\lambda} = \partial_{\mu}v^{\lambda} + \Gamma_{\phantom{a}\mu\nu}^{\lambda}v^{\nu}.
\eea
In addition now the covariant derivative acts on Lorentz indices such that
\bea
\nabla_{\mu}v^{I} = \partial_{\mu}v^{I} + \omega_{\mu\phantom{a}J}^{\phantom{a}I}v^{J}.
\eea
This defines the spin connection $\omega_{\mu\phantom{a}J}^{\phantom{a}I}$.  The spin
connection determines the torsion two-form
\bea
T^{I} = de^{I} + \omega_{\phantom{a}J}^{I}\wedge e^{J} = 0;
\label{rotation}
\eea
the vanishing of the torsion implies the antisymmetry $\omega^{IJ}=-\omega^{JI}$ of the
spin connection.  This with the orthonormality of the tetrad
$e_{\mu}^{\phantom{a}I}e_{\nu I}=\eta_{\mu\nu}$ leads to the metric compatibility condition
$\nabla_{\gamma}g_{\mu\nu}=0$ and determines a unique connection on the manifold called the
affine connection.  With this connection are associated the connection coefficients
$\Gamma_{\phantom{a}\mu\nu}^{\lambda}$.  In terms of these the spin connection coefficients are
$\omega_{\mu\phantom{a}J}^{\phantom{a}I}=\Gamma_{\phantom{a}JK}^{I}e_{\mu}^{\phantom{a}K}$.  This
is general and may be taken as the definition of the $\Gamma$-coefficients.  In a coordinate basis
these are the Christoffel symbols of the second kind.  Now it is clear that the metric compatibility
condition implies that the spin connection is also compatible with the metric:
\bea
\nabla_{\nu}e_{\mu}^{\phantom{a}I}
= \partial_{\nu}e_{\mu}^{\phantom{a}I} - \Gamma_{\phantom{a}\nu\mu}^{\gamma}e_{\gamma}^{\phantom{a}I}
  + \omega_{\nu\phantom{a}J}^{\phantom{a}I}e_{\mu}^{\phantom{a}J} = 0.
\eea
This can be inverted to give back (\ref{rotation}).  The spin connection also determines a curvature
\bea
R_{\mu\nu}^{\phantom{aa}IJ} = \partial_{\mu}\omega_{\nu}^{\phantom{a}IJ}
                              - \partial_{\nu}\omega_{\mu}^{\phantom{a}IJ}
                              + \omega_{\mu}^{\phantom{a}IJ}\omega_{\nu}^{\phantom{a}IJ}
                              - \omega_{\nu}^{\phantom{a}IJ}\omega_{\mu}^{\phantom{a}IJ},
\eea
the components of which determine the curvature two-form
\bea
R_{\phantom{a}J}^{I}
= R_{\mu\nu\phantom{a}J}^{\phantom{aa}I}dx^{\mu}\wedge dx^{\nu}
= d\omega_{\phantom{a}J}^{I} + \omega_{\phantom{a}K}^{I}\wedge\omega_{\phantom{a}J}^{K}.
\label{structure}
\eea
Equations (\ref{rotation}) and (\ref{structure}) are called the Cartan structure equations.  If
the tetrad is invertible then we can define the Ricci tensor
$R_{\mu}^{\phantom{a}I}=R_{\mu\nu}^{\phantom{aa}IJ}e_{\phantom{a}J}^{\nu}$ and Ricci scalar
$R=R_{\mu}^{\phantom{a}I}e_{\phantom{a}I}^{\mu}$.  With this the Hilbert-Palatini action becomes
\bea
S[\bm{e},\bm{\omega}] = \frac{1}{16\pi G_{N}}\int_{\mathcal{M}}d^{4}x
|\mbox{det}(\bm{e})|\left(e_{\phantom{a}I}^{\mu}e_{\phantom{a}J}^{\nu}R_{\mu\nu}^{\phantom{aa}IJ}
- 2\Lambda\right) + S_{matter},
\label{first}
\eea
which up to boundary terms is invariant under local $SO(3,1)$ transformations and local
translations.  The matter action $S_{matter}=S_{matter}[\bm{e},\phi,\psi,\bm{A}]$ is generally
a function of the scalar field $\phi$, spinor field $\psi$ and vector field $\bm{A}$ in addition
to the gravitational field $\bm{e}$.  Variation of the total action with respect to $e^{I}$ gives
the Einstein equations
\bea
e_{\phantom{a}I}^{\mu}R_{\mu\nu}^{\phantom{aa}IJ}
- \frac{1}{2}e_{\phantom{a}M}^{\mu}e_{\phantom{a}N}^{\rho}e_{\nu}^{\phantom{a}J}
  R_{\mu\rho}^{\phantom{aa}MN}
- e_{\nu}^{\phantom{a}J}\Lambda
= 8\pi G_{N}T_{\nu}^{\phantom{a}J}
\eea
as the equations of motion, while variation with respect to $\omega^{IJ}$ gives the metric
compatability condition.  Here the stress-energy tensor
$T_{\mu}^{\phantom{a}I}=\delta S_{matter}/\delta e^{\mu}_{\phantom{a}I}$ defines the
three-form $T^{I}=T_{\mu}^{\phantom{a}I}\epsilon_{\phantom{a}\nu\rho\lambda}^{\mu}
dx^{\nu}\wedge dx^{\rho}\wedge dx^{\lambda}$.

\subsection{Self-Dual Formulation}

We finally come to the self-dual formulation of general relativity
\cite{ashtekara,ashtekarb,ashtekar}, which is the starting point for canonical quantization.
To this end, we consider the integral
\bea
T[\bm{e},\bm{\omega}] = \int_{\mathcal{M}}d^{4}x|\mbox{det}(\bm{e})|
                        e_{\phantom{a}I}^{\mu}e_{\phantom{a}J}^{\nu}\epsilon_{\phantom{aa}MN}^{IJ}
                        R_{\mu\nu}^{\phantom{aa}MN}
\eea
which is a topological term in the sense that it is invariant under local variations
of the tetrad.  An equivalent action for general relativity is therefore
\bea
{}^{+}S[\bm{e},\bm{\omega}]
&=& S[\bm{e},\bm{\omega}] - \frac{i}{32\pi G_{N}}T[\bm{e},\bm{\omega}]\nonumber\\
&=& \frac{1}{16\pi G_{N}}\int_{\mathcal{M}}d^{4}x|\mbox{det}(\bm{e})|
    e_{\phantom{a}I}^{\mu}e_{\phantom{a}J}^{\nu}
    \left[R_{\mu\nu}^{\phantom{aa}IJ}
    - \frac{i}{2}\epsilon_{\phantom{aa}MN}^{IJ}
    R_{\mu\nu}^{\phantom{aa}MN} - 2\Lambda\right].
\eea
Now in terms of the spin connection $\bm{\omega}$ the self-dual
connection is given by
\bea
A_{\mu}^{\phantom{a}IJ}[\bm{\omega}] = \frac{1}{2}\left[\omega_{\mu}^{\phantom{a}IJ}
                                       - \frac{i}{2}\epsilon_{\phantom{aa}MN}^{IJ}
                                       \omega_{\mu}^{\phantom{a}MN}\right],
\eea
and this determines a self-dual curvature
\bea
F_{\mu\nu}^{\phantom{aa}IJ} &=& \partial_{\mu}A_{\nu}^{\phantom{a}IJ}
                              - \partial_{\nu}A_{\mu}^{\phantom{a}IJ}
                              + A_{\mu}^{\phantom{a}IM}A_{\nu M}^{\phantom{aa}J}
                              - A_{\nu}^{\phantom{a}IM}A_{\mu M}^{\phantom{aa}J}\nonumber\\
                            &=&  R_{\mu\nu}^{\phantom{aa}IJ}
                               - \frac{i}{2}\epsilon_{\phantom{aa}MN}^{IJ}
                               R_{\mu\nu}^{\phantom{aa}MN}.
\eea
With this and a change of variable $\bm{\omega}\rightarrow\bm{A}$ the self-dual action becomes
\bea
{}^{+}S[\bm{e},\bm{A}] &=& \frac{1}{8\pi G_{N}}\int_{\mathcal{M}}d^{4}x|\mbox{det}(\bm{e})|
                         (e_{\phantom{a}I}^{\mu}e_{\phantom{a}J}^{\nu}
                         F_{\mu\nu}^{\phantom{aa}IJ} - \Lambda) + S_{matter}.
\eea
Variation of this action with respect to the tetrad field does give the Einstein equations as
can be verified.  Here the connection is complex so the metrics that satisfy the Einstein equations
are complex-valued.  To recover real general relativity certain reality conditions must be imposed
on the phase space.  In particular, we begin with complex general relativity on a real manifold,
and, after defining the Hamiltonian using the self-dual connection, we restrict ourselves to the
real section of the complex phase space.  This is not very straightforward in practice, however,
because it is not obvious that the pull-back of the symplectic structure to the real section is
itself real.  An alternative approach is to start with a real Ashtekar connection for Euclidean
general relativity and later Wick rotate to Lorentzian general relativity.  A more modern approach,
however, is to use a form of Ashtekar variables with a real connection that describes Lorentzian
general relativity, but at the price of having to solve a much more complicated Hamiltonian
constraint.  We will proceed to derive the gravitational Hamiltonian using this Barbero formulation
\cite{barbero} that describes the phase space of Lorentzian general relativity with a real
connection.  Later in Section 5.3 we will go back to the complex self-dual connection and derive
the Kodama state for the simplified Hamiltonian constraint.

\subsection{Classical Phase Space}

The canonical structure is obtained using the usual ADM decomposition (see Appendix C).
Without loss of generality we use the internal gauge symmetry of the gravitational field
to fix the tetrad component $e_{\phantom{a}\hat{I}}^{0}=0$.  Now define on the time-slice
$\Sigma$ the densitized triad
$\tilde{E}_{\phantom{a}\hat{I}}^{i}=\sqrt{q}e_{\phantom{a}\hat{I}}^{i}$ of weight one and
local $SO(3)$ connection
$A_{i}^{\phantom{a}\hat{I}}=\epsilon^{0\hat{I}\hat{J}\hat{K}}A_{i\hat{J}\hat{K}}$,
which are related to the phase space of the induced metric and extrinsic curvature by
\bea
\tilde{E}_{\phantom{a}\hat{I}}^{i}\tilde{E}^{j\hat{I}} = qq^{ij}
\quad
\mbox{and}
\quad
A_{i}^{\phantom{a}\hat{I}} = \Gamma_{i}^{\phantom{a}\hat{I}} + iK_{i}^{\phantom{a}\hat{I}}.
\label{triad}
\eea
Here $\Gamma_{i}^{\phantom{a}\hat{I}}$ is the $SO(3)$ connection of the triad.  From this we
can consider a more general $SO(3)$ connection given by \cite{barbero}
\bea
A_{i}^{\phantom{a}\hat{I}}=\Gamma_{i}^{\phantom{a}\hat{I}}+\gamma K_{i}^{\phantom{a}\hat{I}},
\label{immirzi}
\eea
where $\gamma$ is an arbitrary parameter.  If $\gamma$ is real it is known as the Immirzi
parameter.  It does not play any role in classical dynamics because the extrinsic curvature
term does not show up in the equations of motion unless the action contains fermions.
(It was shown in \cite{perrov} that $\gamma$ appears in the equations of motion as the
coupling constant of a four-point fermion interaction term.)  In terms of the densitized
triad $\tilde{E}_{\phantom{a}\hat{I}}^{i}$ and the connection in (\ref{immirzi}) the action
${}^{+}S$ becomes
\bea
{}^{+}S = \frac{1}{8\pi G_{N}}\int dt\int_{\Sigma}d^{3}x
          \left[iA_{i}^{\phantom{a}\hat{I}}\dot{\tilde{E}}_{\phantom{a}\hat{I}}^{i}
          - iA_{0\hat{I}}\mathcal{G}^{\hat{I}} + iN^{i}\mathcal{V}_{i}
          - \frac{1}{2}\frac{N}{\sqrt{q}}\mathcal{S}\right].
\eea
Here we have seven first class constraints
\bea
\mathcal{G}^{\hat{I}}
&=& D_{i}\tilde{E}^{i\hat{I}} \approx 0\\
\mathcal{V}_{i}
&=& \tilde{E}_{\phantom{a}\hat{I}}^{j}F_{ij}^{\phantom{a}\hat{I}} \approx 0\\
\mathcal{S} &=& \epsilon^{\hat{I}\hat{J}\hat{K}}
                \tilde{E}_{\phantom{a}\hat{I}}^{i}\tilde{E}_{\phantom{a}\hat{J}}^{j}
                \left[F_{ij\hat{K}}
                + \frac{\Lambda}{3}\epsilon_{ijk}
                \tilde{E}_{\phantom{a}\hat{K}}^{k}\right]\nonumber\\
            &\phantom{=}& - \frac{2(1+\gamma^{2})}{\gamma^{2}}
            \tilde{E}_{\phantom{a}\left[\hat{I}\right.}^{i}\tilde{E}_{\phantom{a}\left.\hat{J}\right]}^{j}
            (A_{i}^{\phantom{a}\hat{I}}-\Gamma_{i}^{\phantom{a}\hat{I}})
            (A_{j}^{\phantom{a}\hat{J}}-\Gamma_{j}^{\phantom{a}\hat{J}}) \approx 0,
\label{constraints}
\eea
where $D_{i}$ and $F_{ij\hat{K}}$ are the gauge covariant derivative and curvature of the
connection $A_{i}^{\phantom{a}\hat{I}}$.  (The notation ``$\approx$''
denotes weak equality on the constraint surface.)  In the above action
we identify the term
$iA_{i}^{\phantom{a}\hat{I}}\dot{\tilde{E}}_{\phantom{a}\hat{I}}^{i}$ as the analogue of the
term ``$p_{\mu}\dot{q}^{\mu}$'' in the Legendre transform $\smallint p_{\mu}\dot{q}^{\mu}-H$.
Thus the gravitational (Ashtekar) Hamiltonian is a linear combination of the
constraints
\bea
H_{A} = \frac{1}{8\pi G_{N}}\int dt\int_{\Sigma}d^{3}x
        \left[iA_{0\hat{I}}\mathcal{G}^{\hat{I}} - iN^{i}\mathcal{V}_{i}
        + \frac{1}{2}\frac{N}{\sqrt{q}}\mathcal{S}\right].
\label{hamiltonian}
\eea
The phase space is now described naturally by the $SO(3)$ connection
$A_{i}^{\phantom{a}\hat{I}}$ (gauge potential) and its conjugate momentum
$\tilde{E}_{\phantom{a}\hat{I}}^{i}$ (electric field).  They satisfy the Poisson bracket
relations
\bea
\poisson{A_{i}^{\phantom{a}\hat{I}}}{\tilde{E}_{\phantom{a}\hat{J}}^{j}}
&=& 8\pi\gamma G_{N}\delta_{\hat{J}}^{\hat{I}}\delta_{i}^{j}\tilde{\delta}^{3}(x-x^{\prime})\\
\poisson{A_{i}^{\phantom{a}\hat{I}}}{A_{j}^{\phantom{a}\hat{J}}}
&=& \poisson{\tilde{E}_{\phantom{a}\hat{I}}^{i}}{\tilde{E}_{\phantom{a}\hat{J}}^{j}} = 0.
\eea
Now we know that the first class constraints of a Hamiltonian system generate gauge
transformations.  General relativity is no exception.  The three vector (diffeomorphism)
constraints $\mathcal{V}_{i}$ and scalar (Hamiltonian) constraint $\mathcal{S}$ generate
surface deformations that are equivalent to spacetime diffeomorphisms when the field equations
hold.  The algebra of these surface deformations is a Poisson algebra with structure functions
rather than structure constants.  We have now, in addition, the three ``Gauss law'' constraints
$\mathcal{G}^{\hat{I}}$ that are typical of Yang-Mills theory.  These generate the usual $SO(3)$
gauge transformations.

We make a final note on the canonical theory described here.  The formulation in terms of the
connection $A_{i}^{\phantom{a}\hat{I}}$ turns out to be a generalization of the Einstein theory.
The equations in the metric formulation contain the Ricci scalar which requires the inverse metric
to be defined.  This means that the metric has to be non-degenerate.  The phase space here
is now determined by the set $(A_{i}^{\phantom{a}\hat{I}},\tilde{E}_{\phantom{a}\hat{I}}^{i})$, so
the metric is a derived quantity.  Here all the equations of the Hamiltonian theory are polynomial,
and the inverse triad $\tilde{E}_{i}^{\phantom{a}\hat{I}}$ does not appear in any of them.  The
theory is therefore well-defined even when the triad is degenerate, and therefore the theory admits
solutions with a degenerate metric.  If the triad is restricted to be non-degererate then the
Einstein theory is recovered.

\section{The Gravitational Field: Quantum Theory}

\subsection{Loop Representation}

The phase space of general relativity is now defined and the canonical analysis is complete.
What is left is to quantize the system.  This is done by promoting the constraints
(\ref{constraints}) to operators $\mathcal{G}^{\hat{I}}\rightarrow\hat{\mathcal{G}}^{\hat{I}}$,
$\mathcal{V}_{i}\rightarrow\hat{\mathcal{V}}_{i}$, $\mathcal{S}\rightarrow\hat{\mathcal{S}}$;
these are required to annihilate physical states on the Hilbert space.  In the connection
representation the phase space variables become
\bea
A_{i}^{\phantom{a}\hat{I}} \rightarrow \hat{A}_{i}^{\phantom{a}\hat{I}}
\quad
\mbox{and}
\quad
\tilde{E}_{\phantom{a}\hat{I}}^{i} \rightarrow \hat{\tilde{E}}_{\phantom{a}\hat{I}}^{i}
= -8\pi\gamma G_{N}\frac{\delta}{\delta A_{i}^{\phantom{a}\hat{I}}}.
\eea
Also the Poisson brackets on the phase space are promoted to commutators on the
Hilbert space via $\poisson{x}{y}\rightarrow(1/i\hbar)[\hat{x},\hat{y}]$.  Of course this is
quite difficult in practice due to ordering ambiguities.  General relativity also suffers from
the problem of time, structure functions of the surface deformation algebra etc.  We will not
go into the details here, but refer the reader to the book by Ashtekar \cite{ashtekar}, the
review article by Rovelli \cite{rovelli1} and the book by Rovelli \cite{rovelli2}.

The quantum constraints require that states be gauge invariant functionals of the connection.
And so we are naturally led to consider Wilson loops defined by the self-dual connection.  To
this end we consider an $SU(2)$ spinor description for the canonical variables where
$A_{i}=A_{i\hat{I}}\tau^{\hat{I}}$ and
$\tilde{E}^{i}=2\tilde{E}_{\phantom{a}\hat{I}}^{i}\tau^{\hat{I}}$ in terms of the spin half
$SU(2)$ generators $\tau^{\hat{I}}=-(i/2)\sigma^{\hat{I}}$ in the fundamental representation.
(Here $\sigma^{\hat{I}}$ are the usual Pauli matrices.)  We consider along a continuous piecewise
smooth curve $\map{\gamma}{[0,1]}{\Sigma}$ that is parametrized by $s\rightarrow\{\gamma^{a}(s)\}$
($a\in\{1,2,3\}$), the holonomy of the Ashtekar connection $A_{i}$ such that
\bea
U_{\gamma}(s_{1},s_{2}) = P\mbox{exp}\left[-\int_{s_{1}}^{s_{2}}ds\tdiff{x^{i}(s)}{s}
                          A_{i}^{\phantom{a}\hat{I}}\tau_{\hat{I}}\right].
\eea
This satisfies the differential equation
\bea
\frac{d}{ds}U_{\gamma}(s,s_{0})
+ \tdiff{\gamma^{i}(s)}{s}A_{i}(\gamma(s))U_{\gamma}(s,s_{0}) = 0
\eea
and the boundary condition $U_{\gamma}(s_{0},s_{0})_{A}^{B}=\delta_{A}^{B}$; this simply
says that the vector $U_{\gamma}(s)$ in the vector bundle over the manifold $\mathcal{M}$
is parallel transported along the curve $\gamma$.  Hence the alternate name ``parallel
propagator'' for the holonomy.  Now we define the fundamental set of gauge-invariant loop
operators (for $\Lambda=0$) \cite{rovelli1,rovelli2}.  For a loop $\gamma$ such that
$\gamma^{i}(0)=\gamma^{i}(1)$ and the points $s_{1},s_{2},\ldots,s_{N}\in\gamma$ we have:
\bea
\mathcal{T}[\gamma] &=& -\mbox{Tr}\left[U_{\gamma}(0,1)\right]\\
\mathcal{T}^{i}[\gamma](s)
&=& -\mbox{Tr}\left[U_{\gamma}(s,s)\tilde{E}^{i}(s)\right]\\
\mathcal{T}^{i_{1}i_{2}}[\gamma](s_{1},s_{2})
&=& -\mbox{Tr}\left[U_{\gamma}(s_{1},s_{2})\tilde{E}^{i_{2}}(s_{2})
    U_{\gamma}(s_{2},s_{1})\tilde{E}^{i_{1}}(s_{1})\right]\\
\mathcal{T}^{i_{1}\ldots i_{N}}[\gamma](s_{1},\ldots,s_{N})
&=& -\mbox{Tr}\left[U_{\gamma}(s_{1},s_{N})\tilde{E}^{i_{N}}(s_{N})\ldots
    U_{\gamma}(s_{2},s_{1})\tilde{E}^{i_{1}}(s_{1})\right].
\eea
We will now proceed by introducing some notation for the loop variables that will be needed
below.  Let $\alpha$ and $\beta$ be two loops, and $\ell$ be a segment of a curve.  If
$\alpha$ and $\beta$ intersect at a point $p$ then the loop $\alpha\sharp\beta$ is obtained
when starting at $p$, going first through $\alpha$, then through $\beta$ and finally ending
at $p$.  If a segment of a curve $\ell$ is glued to the loop $\alpha$ then the result is
denoted $\alpha\circ\ell$.  Also the inverse of a loop $\alpha^{-1}$ or of a
segment of a curve $\ell^{-1}$ is just the loop or segment with opposite orientation.  With
this, the loop operators satisfy the properties:
\bea
\mathcal{T}[\alpha] &=& \mathcal{T}[\alpha^{-1}]\\
\mathcal{T}[\alpha] &=& \mathcal{T}[\alpha\circ\ell\circ\ell^{-1}]\\
\mathcal{T}[\alpha]\mathcal{T}[\beta]
&=& \mathcal{T}[\alpha\sharp\beta] + \mathcal{T}[\alpha\sharp\beta^{-1}]\\
\lim_{\epsilon\rightarrow0}\mathcal{T}[\alpha_{x}^{\epsilon}] &=& 2.
\eea
Here $\alpha_{x}^{\epsilon}$ is a loop centered at the point $x$ on the spatial manifold
$\Sigma$ with area $\epsilon$.  Finally we write down the (closed) Poisson algebra of the
lowest-order operators $\mathcal{T}$ and $\mathcal{T}^{i}$:
\bea
\poisson{\mathcal{T}[\alpha]}{\mathcal{T}[\beta]} &=& 0\\
\poisson{\mathcal{T}^{i}[\alpha](s)}{\mathcal{T}[\beta]}
&=& -\frac{i}{2}\Delta^{i}[\beta,\alpha(s)]
    \left[\mathcal{T}[\alpha\sharp_{s}\beta]-\mathcal{T}[\alpha\sharp_{s}\beta^{-1}]\right]\\
\poisson{\mathcal{T}^{i}[\alpha](s)}{\mathcal{T}^{j}[\alpha](t)}
&=& -\frac{i}{2}\Delta^{i}[\beta,\alpha(s)]
    \left[\mathcal{T}^{j}[\alpha\sharp_{s}\beta](t)
    -\mathcal{T}^{j}[\alpha\sharp_{s}\beta^{-1}](t)\right]\\
&\phantom{=}&
    + \frac{i}{2}\Delta^{j}[\alpha,\beta(t)]
    \left[\mathcal{T}^{i}[\beta\sharp_{t}\alpha](s)
    -\mathcal{T}^{i}[\beta\sharp_{t}\alpha^{-1}](s)\right].
\eea
Here we have introduced the notation $\sharp_{s}$ to mean that the breaking and rejoining of
the loops occurs at the intersection of the two loops where the parameter is $s$, and we have
defined the structure functions
\bea
\Delta^{i}[\beta,x] \equiv \oint dt\dot{\beta}^{i}(t)\delta^{3}(\beta(t),x).
\eea
Note that if the loops do not intersect then the Poisson brackets vanish.  These can be obtained
with rigor by using the operator definition for the $\tilde{E}_{\phantom{a}\hat{I}}^{i}$
field.  It is to be understood that these relations define the Poisson brackets and are not the
usual Poisson brackets that are defined by derivations.  The relations above are called the $T$-algebra,
and realizes Isham's proposal \cite{ishkak} of employing a non-canonical Poisson algebra for the
quantization of general relativity.

Let us now prove that loop states solve the constraints of the phase space of general
relativity.  We will do this for any arbitrary loop state.  We have already encountered
a gauge invariant loop, namely the Wilson loop, which is just the trace of the holonomy
of the Ashtekar connection.  The gauge invariance implies that this quantity automatically
solves the Gauss constraint $\mathcal{G}^{\hat{I}}$.  In addition, it is an exact solution
to the scalar constraint $\mathcal{S}$.  So let us consider a general functional of the form
\bea
H[\{\gamma\},\bm{A}] = \prod_{\gamma\in\{\gamma\}}
                       \mbox{Tr}P\mbox{exp}\left[\oint_{\gamma}\bm{A}\right],
\eea
where $\gamma$ is a loop contained in the set of differentiable non-intersecting loops
$\{\gamma\}$.  Then for a functional $\Phi[\bm{A}]$ in the connection representation and
a functional $\Psi[\{\gamma\}]$ in the loop representation, we define the loop transform
\cite{rovsmo1,rovsmo2}
\bea
\Psi[\{\gamma\}] = \int d\mu[\bm{A}]H[\{\gamma\},\bm{A}]\Phi[\bm{A}].
\eea
Here $d\mu[\bm{A}]$ is a measure on the space of connections that is assumed to be
invariant under diffeomorphisms.  The transform is therefore a mapping
$\map{\mathcal{F}}{\Phi}{\Psi[\{\gamma\}]}=\Phi[H[\{\gamma\},\bm{A}]]$.  Operators in the
two representations are related to each other through the transform $\mathcal{F}$ in the
following way.  Let $\hat{\mathcal{O}}$ be an operator acting on the space of connections
and let $\tilde{\mathcal{O}}$ be an operator acting on the space of loops.  Then
$\hat{\mathcal{O}}$ and $\tilde{\mathcal{O}}$ are related through the transformation
\bea
\tilde{\mathcal{O}}\mathcal{F} = \mathcal{F}\hat{\mathcal{O}}^{\dagger}
\eea
provided that $\tilde{\mathcal{O}}H[\{\gamma\},\bm{A}]=\hat{\mathcal{O}}H[\{\gamma\},\bm{A}]$.
Now, the loop transform is the key to understanding how the loop functionals solve all
constraints in the loop representation.  It is known that the Wilson loop satisfies
$\mathcal{G}^{\hat{I}}$ and $\mathcal{S}$.  Therefore the kernel $H[\{\gamma\},\bm{A}]$ also
satisfies $\mathcal{G}^{\hat{I}}$ and $\mathcal{S}$, and it follows that the functional
$\Psi[\{\gamma\}]$ in the loop representation satisfies the constraints for any choice of
functional $\Phi[\bm{A}]$ in the connection representation.  All that remains now is to
impose the vector constraints $\mathcal{V}_{i}$.  However, diffeomorphisms map loops to
other loops in the same knot class, so we conclude that $\Psi[\{\gamma\}]$ will satisfy all
constraints provided that the functional depends on the knot classes of loops.

\subsection{The Spin Network Basis}

The most beautiful result of loop quantum gravity is the discretization of the spatial
manifold.  The non-perturbative quantization of the gravitational field leads to quantum
geometry operators with discrete spectra.  Details on this can be found in
\cite{rovelli2,ashlew1} and references therein.  A self-contained exposition on the
calculation of geometry eigenvalues based on Temperley-Lieb recoupling theory \cite{kaulin}
can be found in \cite{deprov}.  The area operator, for example, has been used to calculate
the Bekenstein-Hawking entropy for an extremal black hole
\cite{rovelli3,krasnov1,krasnov2,abck}.  This result was used to fix the Immirzi
parameter.

Functionals in the loop representation that diagonalize the area and volume operators are
represented by combinatorial structures called spin networks.  A spin network state \cite{deprov}
is a combination of loops called a graph $\bm{\Gamma}\sim\{\gamma\}$, containing a certain number
of edges and vertices.  To each edge we assign a half-integer (spin) $j_{i}$ which labels an
irreducible representation of $SU(2)$, and to each vertex we assign an invariant tensor
$v_{\alpha}$ called an intertwiner in the tensor product of the representations
$j_{1},\ldots,j_{n}$ at which the spins join.  A spin network is then the set
\bea
S = \{\bm{\Gamma},j_{i},v_{\alpha}\}.
\eea
Note that $S$ is an abstract spin network.  The embedding of $S$ into the spatial manifold
is an embedded spin network.  The distinction should be clear from context.  Essentially we
find the holonomy of the connection $\bm{A}$ in the representation $j_{i}$ and use the tensor
$v_{\alpha}$ at each vertex to obtain an invariant functional.  The result is then a state
$\Psi[\bm{A}]$ in the connection representation.  The spin network states are linearly
independent and provide an orthogonal basis for the kinematical Hilbert space
\cite{rovelli1,rovelli2}.  In fact, we can write any state in the connection representation
as a superposition $\ket{\psi}=\sum_{S}\inner{S}{\psi}\ket{S}$.  The coefficient
$\inner{S}{\psi}$ here is the inner product
\bea
\inner{S}{\psi}
= \int_{\overline{\mathcal{A}/\mathcal{G}}}d\mu[\bm{A}]\psi_{S}^{*}[\bm{A}]\psi[\bm{A}],
\eea
where the diffeomorphism invariant measure $d\mu[\bm{A}]$ is over the gauge equivalent classes
of generalized connections $\bar{\mathcal{A}}$.  The configuration space
$\overline{\mathcal{A}/\mathcal{G}}$ is a canonical enlargement of $\mathcal{A}/\mathcal{G}$
\cite{ashish,baez1,ashlew2,ashlew3}.  The enlargement is neccessary if the loop operators are to
remain well defined in the quantum theory where the number of degrees of freedom is infinite.
This measure has only been defined rigorously for $\Lambda=0$.  In the next section we will
discuss the case $\Lambda\neq0$.  There, we will see how the loop transform is related to
Witten's partition function when we discuss the Kodama wavefunction, and obtain a known
topological invariant as an exact quantum state in the loop representation.  As we will see,
the presence of $\Lambda$ can be incorporated into the measure of Chern-Simons theory.

\subsection{An Exact State}

Recall the scalar constraint $\mathcal{S}$ in (\ref{constraints}).  If we set $\gamma=i$ as for the
self-dual connection, then the last term vanishes and we get
\bea
\mathcal{S} &=& \epsilon^{\hat{I}\hat{J}\hat{K}}
                \tilde{E}_{\phantom{a}\hat{I}}^{i}\tilde{E}_{\phantom{a}\hat{J}}^{j}
                \left[F_{ij\hat{K}}
                + \frac{\Lambda}{3}\epsilon_{ijk}
                \tilde{E}_{\phantom{a}\hat{K}}^{k}\right] \approx 0.
\eea
This is the Hamiltonian constraint in the original self-dual formulation.  It is clear that
a solution to this $\mathcal{S}$ (and also to $\mathcal{G}^{\hat{I}}$ and $\mathcal{V}_{i}$)
will have to satisfy
\bea
F_{ij\hat{K}} = -\frac{\Lambda}{3}\epsilon_{ijk}\tilde{E}_{\phantom{a}\hat{K}}^{k}.
\label{curv}
\eea
If $A_{i\hat{I}}=i\sqrt{\Lambda/3}f\delta_{i\hat{I}}$ is chosen for the connection
$A_{i}^{\phantom{a}\hat{I}}$ with arbitrary function $f=f(t)$, then the curvature becomes
$F_{ij\hat{K}} = -\frac{\Lambda}{3}f^{2}\epsilon_{ij\hat{K}}$.  Now the connection defined
in (\ref{triad}) is purely imaginary so the $SO(3)$ connection
$\Gamma_{i}^{\phantom{a}\hat{I}}$ of the triad vanishes which means that $\Sigma$ is a flat
manifold.  The canonical momentum becomes
$\tilde{E}_{\phantom{a}\hat{I}}^{i} = f^{2}\delta_{\phantom{a}\hat{I}}^{i}$ so that the
spatial metric is $q_{ij}=f^{2}\delta_{ij}$.  To find the evolution equation for $f$ we
write down the Hamilton equations of motion for the canonical variables:
\bea
\dot{A}_{i}^{\phantom{a}\hat{I}} = \poisson{A_{i}^{\phantom{a}\hat{I}}}{H_{A}}
\quad
\mbox{and}
\quad
\dot{\tilde{E}}_{\phantom{a}\hat{I}}^{i} = \poisson{\tilde{E}_{\phantom{a}\hat{I}}^{i}}{H_{A}}.
\eea
For simplicity we take the shift vector $N^{i}$ in (\ref{hamiltonian}) to be zero.  In other words,
the spacetime metric has no $t$-$x^{i}$ cross terms.  The evolution for $f$ is then given by
$\dot{f}=N\sqrt{\Lambda/3}f^{4}$, and taking the lapse function $N=1/\sqrt{q}=1/f^{3}$ leads to
$\dot{f}=\sqrt{\Lambda/3}f$ or $f=\mbox{exp}[\sqrt{\Lambda/3}t]$.  So the line element is
\bea
ds^{2} = -dt^{2} + q_{ij}dx^{i}dx^{j}
       = -dt^{2} + \mbox{exp}\left[2\sqrt{\frac{\Lambda}{3}}t\right](dx^{i})^{2},
\eea
which is immediately recognized as de Sitter spacetime.  This is also a Friedmann-Robertson-Walker
spacetime with flat spatial manifold and scale factor $R(t)=\mbox{exp}[\sqrt{\Lambda/3}t]$.  It is
the only self-dual solution to the Einstein equations for Lorentzian manifolds.  Now, to see how
this solution is related to topological field theory in $\Sigma$, we take $S(\bm{A})$ to be a
Hamilton-Jacobi functional on the space of connections $\mathcal{A}$ in $\Sigma$.  This implies
that the canonical momentum is
\bea
\tilde{E}_{\phantom{a}\hat{I}}^{i} = -8\pi G_{N}\fdiff{S(\bm{A})}{A_{i}^{\phantom{a}\hat{I}}},
\eea
but from the curvature (\ref{curv}) we have then
\bea
F_{jk\hat{I}} = \frac{8\pi\Lambda G_{N}}{3}\epsilon_{ijk}\fdiff{S(\bm{A})}{A_{i}^{\phantom{a}\hat{I}}}.
\eea
In terms of $A_{i}^{\phantom{a}\hat{I}}$ the curvature is
\bea
F_{ij}^{\phantom{aa}\hat{I}}
= \partial_{i}A_{j}^{\phantom{a}\hat{I}} - \partial_{j}A_{i}^{\phantom{a}\hat{I}}
  + \epsilon_{ab}^{\phantom{aa}\hat{I}}A_{i}^{a}A_{j}^{b}.
\eea
Integration then gives the Hamilton-Jacobi functional
\bea
S(\bm{A}) = \frac{3}{8\pi\Lambda G_{N}}\int_{\Sigma}Y_{CS},
\quad
Y_{CS} = \frac{1}{2}\mbox{Tr}\left[\bm{A}\wedge d\bm{A} + \frac{2}{3}\bm{A}\wedge\bm{A}\wedge\bm{A}\right].
\label{kcs}
\eea
From this we are led to the semiclassical state
\bea
\Psi_{K}(\bm{A}) = \mathcal{N}\mbox{exp}\left[\frac{3}{8\pi\Lambda G_{N}}\int_{\Sigma}Y_{CS}\right]
\eea
which describes de Sitter spacetime.   This state is, however, an exact state in quantum gravity.
It is annihilated by all the constraints:
\bea
\hat{\mathcal{G}}^{\hat{I}}\Psi_{K}(\bm{A}) = 0,
\quad
\hat{\mathcal{V}}_{i}\Psi_{K}(\bm{A}) = 0
\quad
\mbox{and}
\quad
\hat{\mathcal{S}}\Psi_{K}(\bm{A}) = 0.
\eea
This means that $\Psi_{K}(\bm{A})$ is a physical state of non-perturbative quantum gravity with a
non-zero cosmological constant \cite{smolin1}.  This is the Kodama state, whose classical limit is
de Sitter spacetime.  States of the form
\bea
\Psi[\bm{A}] = \mathcal{N}\mbox{exp}\left[\frac{3}{8\pi\Lambda G_{N}}
               \int_{\Sigma}(Y_{CS}+8\pi\Lambda G_{N}S^{\prime}[\bm{A}])\right]
\eea
can then be used to obtain linearized gravity on de Sitter spacetime, i.e. gravitons propagating over
a classical background.  The normalization factor $\mathcal{N}$ satisfies
$\fsent{\mathcal{N}}{\bm{A}}=0$, and depends on the topology of the manifold.  In fact, Soo \cite{soo}
pointed out that a more general solution would sum over all inequivalent topologies of $\Sigma$ so that
\bea
\Psi[\bm{A}] = \sum_{\{Top(\Sigma)|\partial\Sigma=0\}}\mathcal{N}_{\Sigma}
               \mbox{exp}\left[\frac{3}{8\pi\Lambda G_{N}}\int_{\Sigma}Y_{CS}\right].
\eea
The boundary of $\Sigma$ is required to vanish so that $\Psi[\bm{A}]$ is still annihilated by
all the constraints.  This sum, however, is very large and in fact unknown, because a complete
classification of three-manifolds is lacking at this time.

Finally we come to a very striking property of the Kodama state, which serves as a point of
convergence for all the material that has been discussed up to this point.  So let us take the
loop transform of the Kodama state.  Allowing for either Euclidean or Lorentzian manifold $\mathcal{M}$
by introducing the integer $\epsilon=\{i,1\}$, we have
\bea
\Psi[\{\gamma\}] = \int d\mu[\bm{A}]H[\{\gamma\},\bm{A}]
                   \mathcal{N}\mbox{exp}\left[\frac{3\epsilon}{8\pi\Lambda G_{N}}Y_{CS}\right].
\label{transformkodama}
\eea
We want to be able to evaluate the loop transform for states with $\Lambda\neq0$ so as to include
the Kodama state.  The above integral, however, is undefined unless the loops are framed.  With
this framing regularization the group $SU(2)$ of spin networks becomes ``quantum deformed'' to
$SU_{q}(2)$ with deformation parameter
\bea
q = \mbox{exp}\left[\frac{2\pi}{k+2}\right],
\quad
\mbox{with}
\quad
k = \frac{6\pi}{G_{N}^{2}\Lambda}.
\eea
Clearly then $SU_{q}(2)\rightarrow SU(2)$ for $\Lambda\rightarrow0$, $q\rightarrow1$.  For
Euclidean manifold $\epsilon=i$ this is exactly Witten's partition function with gauge group
$G=SU(2)$ and the Wilson loops all in the same two-dimensional fundamental representation of
$SU(2)$.  In loop space the Euclideanized Kodama state is then the bracket polynomial in the
variable $\Lambda$.  Unfortunately, this picture is a bit more difficult for the Lorentzian
case $\epsilon=1$ because now the integral (\ref{transformkodama}) has to be taken over a
contour since the connection is complex.  It is possible to analytically continue the
deformation parameter to complex values of the Chern-Simons level.  With the loops now framed,
the spin networks become essentially blown up to a two-dimensional surface
\cite{smolin2,bms,majsmo}.  Edges become tubes with ruling.  The ruling means that the original
loop before deformation can be identified.  Vertices become spheres with punctures, each puncture
being the location where a tube intersects the sphere.  The punctures are now labelled by
intertwiners of the quantum group $G_{q}=SU_{q}(2)$.  The intertwiners depend on the level of
the Chern-Simons theory, and are none other than the space of conformal blocks.  The reader will
find this easy to visualize if he/she thinks of the spin networks as just trivalent graphs and
replaces the edges by tubes and the vertices by junctures of tubes that form non-singular
surfaces.  We refer the reader to the literature cited in this subsection for the handling of
framing with respect to the networks and the surfaces.

Now the picture is complete.  Functionals of non-perturbative quantum gravity with a positive
cosmological constant in the loop representation are finite-genus ruled Riemann surfaces
with marked points.  It is remarkable that the structures which naturally describe
perturbative string theory \cite{gsw,hatfield} have emerged as functionals of a non-perturbative
quantization of the gravitational field.  One thing still lacking in perturbative string theory
is a background-independent formulation called M(ystery) theory.  An ambitious program launched
by Smolin \cite{marsmo1,marsmo2,smolin3,smolin4} is to describe M theory as a deformation of
abstract spin networks.

\section{Outlook}

In this review we discussed in detail non-Abelian Chern-Simons theory for arbitrary gauge
group on a three-dimensional manifold with boundary.  We examined the relation of this field
theory to the WZW conformal field theory in two dimensions and saw that, after canonical
quantization, the wavefunctions of the Chern-Simons theory are the conformal blocks of the
WZW current algebra at level $k$.  By using the Chern-Simons action as the measure of a path
integral which defines the expectation value of the Wilson loop, we described Witten's
intrinsic three-dimensional definition for the Jones polynomial.  We then described a
reformulation of general relativity on an $SO(3)$ Yang-Mills phase space, and defined the
fundamental loop representation of the non-canonical $T$-algebra for $\Lambda=0$ which led
to a purely combinatorial structure called the spin network.  For $\Lambda\neq0$ we solved
the constraints to obtain an exact state of quantum gravity whose classical limit is de
Sitter spacetime.  The presence of $\Lambda$ led to the framing of loops and therefore to a
quantum deformation of the spin networks, resulting in a discrete spacetime of punctured
Riemann surfaces with finite genus.

In loop quantum gravity, we have understood that via the loop transform one can represent states
of quantum gravity via Wilson loops (and integrals of Wilson loops over the underlying gauge field
$\bm{A}$), and hence by the geometry of knots and links embedded in the three-space.  The loop
transform associates a functional integral to a given loop in three-space, and it is seen that in
the case of the Chern-Simons state that this transform is, up to framing, a topological invariant
of the loop (as discussed in more detail below).  One knows that these same topological invariants
arise by taking bracket evaluations of the knots at appropriate values of the bracket variable $A$.
As a result, it is a well-grounded move to shift from the functional integral approach to an
approach that evaluates knotted and embedded spin networks in three-dimensional space.  Each such
evaluation can be regarded as a specific evaluation of a quantum gravity loop transform, and one
can look for geometric and physical intepretations of these evaluations.  One can think of each
loop state as having all its geometry concentrated along the loop in the fashion of a
three-dimensional embedded delta function.  The metric that arises here assigns an area measurement
to any test two-dimensional surface that encounters the spin network loop.  Each intersection point
gives the value $8\pi\gamma\hbar G_{N}c^{-3}\sqrt{j(j+1)}$, where $j$ labels the spin network edge,
and we have retained the speed of light $c$ and Planck's constant $\hbar$.  One adds up all the
contributions to the area coming from the different spin network lines in a given or chosen family
of embedded spin networks in three-dimensional space. The remarkable point about this method is
that in the context of loop quantum gravity, the area measurements are quantized in this fashion.
The fundamental quantum of area is given by a surface with a single puncture, with the edge carrying
a spin $j=1/2$.  In this case
$A=4\sqrt{3}\pi\gamma\hbar G_{N}c^{-3}\approx1.808\times10^{-69}$ $\mbox{m}^{2}$.  Surfaces make
quantum jumps between states in the spectrum of the area operator.  This, in turn reflects on many
other issues such as the discrete nature of space and time at the Planck scale and the information
content of black holes.  For more information on these and related matters we refer the reader
to \cite{rovelli3,krasnov1,krasnov2,abck,ars,rovsmo3,ashlew4,krasnov3,major}.

Having discussed the achievements of the approach to quantum gravity presented in this review, we
will now turn to a brief discussion of several key problems that are still open.  These include:
\begin{itemize}

\item
\emph{The physical Hilbert space}: Time evolution is governed by the Wheeler-de Witt equation
$\hat{S}\psi=0$, where $\hat{S}$ is the operator corresponding to the complicated Hamiltonian
constraint in (\ref{constraints}).  The theory should express four-diffeomorphism invariance
but the canonical decomposition requires a choice of external time parameter, thus breaking
general covariance.  The states described here form the ``kinematical'' Hilbert space; that
is, these states only solve the quantum constraints $\mathcal{G}^{\hat{I}}$ and
$\mathcal{V}_{i}$.  What is needed is a set of functionals that will solve the Wheeler-de Witt
equation and define the physical Hilbert space, and on it a physical inner product.  A possible
candidate for the Wheeler-de Witt operator has been defined by Thiemann in \cite{thiemann1},
but it is not obvious that this operator in the classical limit describes general relativity
at all.

\item
\emph{Reconstruction problem}: Any candidate theory of quantum gravity should in some limit
reproduce a classical background metric that satisfies the Einstein equations.  Furthermore,
we expect that a good theory of quantum gravity coupled to matter should in some low-energy
limit reproduce quantum field theory in curved spacetime.  This is, however, a very non-trivial
problem to address.  Matter has been coupled to gravity via the self-dual variables and has
been quantized \cite{rovmor1,rovmor2,baekra,thiemann2,monrov}.  The problem has only been
solved on the kinematical Hilbert space.  As mentioned above, an inner product for physical
states is currently lacking in loop quantum gravity.  Therefore the dynamics of matter coupled
to gravity, as for example transition amplitudes to describe scattering processes within this
framework is currently still out of reach.

\item
\emph{Normalization of the Kodama wavefunction}: An analogous wavefunction to the Kodama state
exists for $SU(2)$ Yang-Mills theory given by
\bea
\Psi(\bm{A}) = \mbox{exp}\left[\frac{1}{g^{2}}\int_{\Sigma}Y_{CS}\right],
\label{ymstate}
\eea
where $g$ is the coupling constant and $Y_{CS}$ is given in (\ref{kcs}).  This state is a solution
to $[E_{i}^{a}-iB_{i}^{a}]\Psi(\bm{A})=0$ with $E_{i}^{a}=ig^{2}\delta/\delta A_{a}^{i}$.  Witten
\cite{witten3} pointed out that the physical norm of this theory, which is given by
\bea
\langle\Psi|\Psi\rangle = \int dA|\Psi(\bm{A})|^{2}
                        = \int dA\mbox{exp}\left[\frac{2}{g^{2}}\int_{\Sigma}Y_{CS}\right],
\eea
is infinite because the configuration space is unbounded from below.  He conjectured that this
should be true also for the Kodama state.  This issue is addressed in \cite{fresmo}, where it is
argued that this argument does not extend to gravity because the Kodama state, as any other states
that are physical states, are expected to be nonnormalizable in the kinematical inner product.
The Yang-Mills theory above is not diffeomorphism-invariant and hence the state (\ref{ymstate}) is
nonnormalizable in the physical inner product.  In \cite{fresmo}, in order to address the issue of
normalizability in the physical inner product, the authors studied the linearized Kodama state over
a de Sitter background.  They found that although this linearized state is delta-function
normalizable in the Euclidean theory, it is not normalizable in the Lorentzian theory.  This raises
an important issue as to whether the full Kodama state discussed in this paper is a physically
reasonable state to describe the ground state of quantum gravity with a positive cosmological
constant.

\end{itemize}

Most of these issues may be solved in the future using the state sum and spin foam models
for quantum gravity that are currently under intensive investigation.  These are based on a
covariant path integral approach and hence do not break general covariance.  There is not
room in this short review to do more than mention these aspects.  We refer the reader to
\cite{barcra1,barcra2,crane1,cpr,crane2,crayet1,crayet2,acs,crane3,chrcra} for more
information on state sum models and \cite{reisenberger,baez2,frekra,oriti,perez1} for
more information on spin foam models.  Several important problems have now been addressed
in this direction:
\begin{itemize}

\item
The reconstruction problem has begun in \cite{markopoulou1,markopoulou2}, employing
renormalization group methods of statistical physics to find the low-energy limit of
spin foam models.

\item
Work is in progress on spin foam models of matter coupled to quantum gravity
\cite{mikovic1,mikovic2,oripfe}; the question of particle scattering within the framework
of background-independent quantum gravity has been addressed in \cite{modrov}; an attempt
has been made at obtaining the graviton propagator via spin foams in \cite{rovelli4}.

\item
Recently the idea has been put forward, that perturbative quantum gravity can be treated as
an expansion around topological $BF$ theory \cite{fresta}.  This is related to the fact that
gravity can be written as $BF$ theory with the constraint that the $B$ field comes from the
metric.  An implementation of this idea has been applied to standard Yang-Mills theory in
flat spacetime \cite{rovspe} and was shown to be a viable technique for perturbation theory
of generally covariant theories.

\end{itemize}

We conclude with a few remarks on the background dependence of matter quantum field theories.
There it is always assumed from the outset that the background spacetime is continuous to
arbitrarily small scales.  In the non-perturbative quantization of pure gravity we generically
end up with a discrete space with length, area and volume eigenvalues.  So the initial
assumption of quantum field theory is false in this context, and may in fact be the origin
of the ultraviolet divergences that plague the quantum field theories.  Indeed in his QSD
V article, Thiemann \cite{thiemann3} shows that the divergences can be fully accounted for
if we neglect the discreteness of space, and has argued that the Planck scale may provide
a natural cutoff regularization of the Hamiltonian constraint of matter quantum field
theories.  We can go further.  By looking at the modifications to matter interactions
after taking the discreteness of space into account, we may find a better understanding
of nature, built on a well-defined set of principles.

\section*{Acknowledgments}

This review is an expanded version of a report that was written for a graduate course on knots
and quantum gravity, taken by T.L. during the summer of 2004 at the University of Waterloo.  T.L.
is supported by the Natural Sciences and Engineering Research Council of Canada under grant 101203.

It gives L.H.K. pleasure to thank the National Science Foundation for support of this
research under NSF Grant DMS-0245588. Much of his effort was sponsored by the Defense Advanced
Research Projects Agency (DARPA) and Air Force Research Laboratory, Air Force Materiel Command,
USAF, under agreement F30602-01-2-05022.  The U.S. Government is authorized to reproduce and
distribute reprints for Government purposes notwithstanding any copyright annotations thereon.
The views and conclusions contained herein are those of the authors and should not be interpreted
as necessarily representing the official policies or endorsements, either expressed or implied,
of the Defense Advanced Research Projects Agency, the Air Force Research Laboratory, or the U.S.
Government. (Copyright 2005.)

The authors would both like to thank the University of Waterloo and the Perimeter Institute for
their hospitality during much of the work on this paper.

The authors would also like to thank Kirill Krasnov for comments, and the referees for pointing
out several errors in an earlier draft of the manuscript, and for their suggestions which
improved the presentation of the review.

\appendix

\section{Conventions}

Here we collect the conventions used in the paper.  Unless otherwise stated, we work on a
four-dimensional Lorentzian manifold $\mathcal{M}$ which admits constant time hypersurfaces
$\Sigma$.  We use the following index conventions: lower-case greek letters
$\mu,\nu,\ldots\in\{0,1,2,3\}$ are spacetime indices; lower-case roman letters
$i,j,\ldots\in\{1,2,3\}$ are spatial indices; upper-case roman letters
$I,J,\ldots\in\{0,1,2,3\}$ are internal $SO(3)$ tangent space indices; upper-case roman
letters with hats $\hat{I},\hat{J},\ldots\in\{1,2,3\}$ are gauge-fixed internal $SO(3)$
tangent space indices.  Spacetime indices are raised and lowered with the spacetime metric
$g_{\mu\nu}$, spatial indices are raised and lowered with the induced spatial metric $q_{ij}$
on a constant time slice, and internal indices are raised and lowered with the Minkowski metric
$\eta_{IJ}=\mbox{diag}(-1,+1,+1,+1)$.  Here spacetime and spatial indices are consistently
written before the tangent space indices for all quantities on $\mathcal{M}$.  The Einstein
summation convention is used for repeated upper and lower indices.  Units are employed for
which the speed of light $c$ and Planck's constant $\hbar$ are unity, and Newton's constant
$G_{N}$ is retained.

\section{Hamiltonian Formulation of Classical Mechanics}

For convenience we summarize here the Hamiltonian formulation of a classical unconstrained
system with a finite number of degrees of freedom \cite{arnold,gps}.  We consider
a system that is governed by the action $S[q,\dot{q};t]=\int L(q,\dot{q};t)dt$, where $t$ is
an integration parameter with an overdot denoting $d/dt$, and $L$ is known as the system's
Lagrangian.  This is a functional, and is defined as the kinetic energy minus the potential
energy.  In most cases the kinetic energy $T=T(\dot{q})$ is a function of the system's velocities
$\dot{q}=\dot{q}(t)=\{\dot{q}^{\alpha}(t)\}_{\alpha=1}^{n}$ and the potential energy $V=V(q)$ is
a function of the system's coordinates $q=q(t)=\{q^{\alpha}(t)\}_{\alpha=1}^{n}$.  The set of
all coordinates $\{q^{\alpha}\}_{\alpha=1}^{n}$ are said to define an $n$ dimensional manifold
called the configuration space $\mathcal{Q}$.  However, the defining action $S[q,\dot{q};t]$
exludes the more interesting gauge theories because the Lagrangian only describes systems with
a finite number of degrees of freedom.  For the former we would need to consider
$n\rightarrow\infty$.  Furthermore, the parameter $t$ in most nonrelativistic systems is the
time which measures evolution, but this breaks down when we consider relativistic motion since
time is no longer a parameter.

Now, the functional $S=S[q,\dot{q};t]$ is differentiable, and the equations of motion of the system
are obtained by demanding that the action be stationary with respect to variations of $q$ and $\dot{q}$.
We find that
\bea
\delta S = \int dt\left(\pdiff{L}{q^{\alpha}}\delta q^{\alpha}
           + \pdiff{L}{\dot{q}^{\alpha}}\delta\dot{q}^{\alpha}\right)
         = \int dt\left(\pdiff{L}{q^{\alpha}}
           - \frac{d}{dt}\pdiff{L}{\dot{q}^{\alpha}}\right)\delta q^{\alpha}.
\label{vderivative}
\eea
Here we integrated by parts using $\delta\dot{q}^{\alpha}=\tsent{(\delta q^{\alpha})}{t}$.
The boundary term vanishes because we are taking the variation with respect to
fixed endpoints.  If we impose $\delta S=0$ for arbitrary $\delta q^{\alpha}$,
then it follows that
\bea
\pdiff{L}{q^{\alpha}} - \frac{d}{dt}\pdiff{L}{\dot{q}^{\alpha}} = 0.
\label{ele1}
\eea
This is a set of $n$ second order differential equations in $q^{\alpha}$,
$\dot{q}^{\alpha}$, and $t$.  These are called the Euler-Lagrange equations
of motion, and by Hamilton's principle of least action (which states that the
motion of a mechanical system coincides with the extremals of the functional
$S[q,\dot{q};t]$) determine the (time) evolution of the system.  Furthermore, the
canonical (generalized) momenta of the system are given by
\bea
p_{\alpha} = \pdiff{L}{\dot{q}^{\alpha}};
\label{momenta}
\eea
these are of central importance in the Hamiltonian formulation to be introduced below.
They are generalized momenta in the sense that they represent a generalization of the
Newtonian notion of momenta given by $\bm{p}=m\dot{\bm{q}}$.  Equations (\ref{ele1})
can be shown to reproduce Newton's second law $\bm{F}=m\bm{a}$ in the form
$\tsent{\bm{p}}{t}=-\nabla V$.

An equivalent approach to Lagrangian mechanics outlined above is to use a Legendre
transform to obtain a set of first order differential equations to replace (\ref{ele1}).
To this end, we let $L=L(q,\dot{q};t)$ and $\dot{p}_{\alpha}=\psent{L}{q^{\alpha}}$.
Now, we define the Hamiltonian function $H=H(q,p;t)$ through the Legendre transform
\bea
H(q,p;t) = \dot{q}^{\alpha}p_{\alpha} - L(q,\dot{q};t).
\label{legendre}
\eea
The differential of this function is
\bea
dH &=& \dot{q}^{\alpha}dp_{\alpha} + p_{\alpha}d\dot{q}^{\alpha}
       - \pdiff{L}{q^{\alpha}}dq^{\alpha}
       - \pdiff{L}{\dot{q}^{\alpha}}d\dot{q}^{\alpha} - \pdiff{L}{t}dt\nonumber\\
   &=& \dot{q}^{\alpha}dp_{\alpha} - \dot{p}_{\alpha}dq^{\alpha} - \pdiff{L}{t}dt.
\label{hdiff1}
\eea
In general, however, we should expect the Hamiltonian $H=H(q,p;t)$ to have
differential of the form
\bea
dH = \pdiff{H}{q^{\alpha}}dq^{\alpha} + \pdiff{H}{p_{\alpha}}dp_{\alpha} + \pdiff{H}{t}dt.
\label{hdiff2}
\eea
Comparing this to (\ref{hdiff1}) we obtain the Hamilton equations of motion
\bea
\dot{q}^{\alpha} = \pdiff{H}{p_{\alpha}}
\quad
\mbox{and}
\quad
\dot{p}_{\alpha} = -\pdiff{H}{q^{\alpha}},
\label{hamilton1}
\eea
for which the Lagrangian must satisfy $\psent{L}{t}=-\psent{H}{t}$.  If the transformations
connecting the rectangular and canonical coordinates are independent of time so that the
kinetic energy is a homogeneous quadratic function of the $\dot{q}^{\alpha}$, and the
potential energy is independent of velocity, then it follows that the Hamiltonian is equal
to the total energy of the system and is a conserved quantity provided it does not
explicitly depend on time.  In fact, the Hamiltonian is a conserved quantity provided that
the system is closed, i.e. a system which does not interact with its surroundings.

Let's now consider a function denoted by $A=A(q,p;t)$, for which we calculate the total
time derivative to be
\bea
\tdiff{A}{t} &=&  \left(\pdiff{A}{q^{\alpha}}\tdiff{q^{\alpha}}{t}
                  + \pdiff{A}{p_{\alpha}}\tdiff{p_{\alpha}}{t}\right) + \pdiff{A}{t}\nonumber\\
             &=&  \left(\pdiff{A}{q^{\alpha}}\pdiff{H}{p_{\alpha}}
                 - \pdiff{A}{p_{\alpha}}\pdiff{H}{q^{\alpha}}\right) + \pdiff{A}{t}.
\eea
Here we made use of the Hamilton equations (\ref{hamilton1}).  This can be written
more abstractly if we define the Poisson bracket\foot{The Poisson bracket satisfies
the following identities:
\begin{enumerate}
\item
$\poisson{F}{G}=-\poisson{G}{F}$
\item
$\poisson{F+H}{G}=\poisson{F}{G}+\poisson{H}{G}$
\item
$\poisson{FH}{G}=F\poisson{H}{G}+\poisson{F}{G}H$
\item
$\poisson{F}{\poisson{G}{H}}+\poisson{G}{\poisson{H}{F}}+\poisson{H}{\poisson{F}{G}}=0$
\end{enumerate}}
\bea
\poisson{A}{B} = \left(\pdiff{A}{q^{\alpha}}\pdiff{B}{p_{\alpha}}
                 - \pdiff{A}{p_{\alpha}}\pdiff{B}{q^{\alpha}}\right),
\label{poisson}
\eea
so that
\bea
\tdiff{A}{t} = \poisson{A}{H} + \pdiff{A}{t}.
\label{hamilton2}
\eea
In most cases $A$ will contain time dependence implicitly so that $\psent{A}{t}=0$, and
thus we have
\bea
\tdiff{A}{t} = \poisson{A}{H}.
\label{hamilton3}
\eea
This beautiful expression contains in it all of classical mechanics.  If we
evaluate the Poisson bracket for $q^{\alpha}$ and $p_{\alpha}$ we obtain
the equations (\ref{hamilton1}).  In this form, we see that the Hamiltonian
is the generator of time translations.  The set of all coordinates
$\{q^{\alpha},p_{\alpha}\}_{\alpha=1}^{n}$ are said to define a $2n$ dimensional
manifold called the phase space $\Gamma$.  A point $(q^{\alpha},p_{\alpha})$ on
$\Gamma$ is called a state, and a function $F=F(q,p;t)$ on $\Gamma$ that satisfies
Hamilton's equation $\dot{F}=\poisson{F}{H}$ is called an observable.  The set of
$n$ second order equations on $\mathcal{Q}$ have been replaced by a set of $2n$
first order equations on $\Gamma$.

A more absract formulation can be obtained if we define the phase space as a dual
tangent bundle, or cotangent bundle, with coordinates $(q^{\alpha},p_{\alpha})$ on
which the Hamiltonian $\map{H}{T^{*}(N)}{\mathbb{R}}$ is defined.  This can be seen
in the following way.  Recall that the Lagrangian $L=L(q,\dot{q};t)$ is defined over
an $n$-dimensional manifold called the configuration space $\mathcal{Q}$.  The
$\dot{q}^{\alpha}$ are the components of a vector at the point $(q^{1},\ldots,q^{n})$.
Thus the Lagrangian $\map{L}{T(N)}{\mathbb{R}}$ is a function on the tangent bundle.
Now the Hamiltonian is obtained from the Legendre transform (\ref{legendre}).  The
first term $\dot{q}^{\alpha}p_{\alpha}$ can be visualized as a pairing of an element
of the tangent space with its dual.  In replacing $\dot{q}^{\alpha}$ with $p_{\alpha}$
as the new set of independent variables, the Legendre transform replaces the tangent
bundle $T(N)$ with the cotangent bundle $T^{*}(N)$ and thus defines the Hamiltonian as
a function $\map{H}{T^{*}(N)}{\mathbb{R}}$.  This explains why in the Einstein summation
$q^{\alpha}p_{\alpha}\equiv\sum_{\alpha}q_{\alpha}p_{\alpha}$ for some canonical set of
variables $\{q^{\alpha},p_{\alpha}\}_{\alpha=1}^{n}$ we write the position $q^{\alpha}$
with contravariant index and the momentum $p_{\alpha}$ with covariant index.

The connection between conserved quantities and the symmetries of a Lagrangian is contained
in Noether's theorem.  The essence of this theorem is that, if the Lagrangian is unaffected
by a transformation that alters one of its coordinates, then the Lagrangian is invariant,
or symmetric, under the given transformation, and the corresponding current is then
conserved.  In particular, there are seven constants associated with the motion of a closed
system: homogeneity of time implies that the Lagrangian is invariant under time translations,
which leads to the conservation of energy; homogeneity of space implies that the Lagrangian
is invariant under spatial translations, which leads to the conservation of linear momentum;
isotropy of space implies that the Lagrangian is invariant under spatial rotations, which
leads to the conservation of angular momentum.  In the Hamiltonian formulation, a conserved
quantity is simply one for which the Poisson bracket vanishes, i.e.
$\tsent{A}{t} = \poisson{A}{H} = 0$.  This provides a simple proof of the conservation of
energy since $\poisson{H}{H}=0$ must be satisfied by the antisymmetry property of the Poisson
bracket.  Finally, we define a cyclic coordinate $\tilde{q}^{\alpha}$ to be one that does not
appear explicitly in the Lagrangian.  Then (\ref{ele1}) reduces to $\dot{p}_{\alpha}=0$, in
which case the Hamiltonian does not explicitly depend on $\tilde{q}^{\alpha}$ either.  Thus
the canonical momentum conjugate to a cyclic coordinate is conserved, i.e.
$\poisson{p_{\alpha}}{H}=0$.

\section{ADM Decomposition}

In order to cast the Einstein-Hilbert action $S[g]=(1/16\pi G_{N})\smallint d^{4}x\sqrt{g}(R-2\Lambda)$
with $R=R[g]$ into Hamiltonian form, we need a time with which we can define the conjugate momenta.
To this end, we split the spacetime into a space and a time.  We assume that $\mathcal{M}$ has the
topology $[0,1]\times\Sigma$ where $\Sigma$ is an open or closed three-dimensional surface; this is
a segment of a universe between initial surface $\{0\}\times\Sigma$ and final surface
$\{1\}\times\Sigma$ both assumed to be spacelike.  We also assume the existence of constant time
hypersurfaces $\{\Sigma_{t}\}$ endowed with coordinate systems $\{x^{i}\}$ and induced metrics
$\{q_{ij}(t,x^{k})\}$.  After an infinitesimal time translation from $\Sigma_{t}$ to
$\Sigma_{t+dt}$, the change in proper time will be $d\tau=Ndt$, and this leads in general to a
spatial shift $x^{i}(t+dt)=x^{i}(t)-N^{i}dt$.  Here $N=N(t,x^{i})$ is the lapse function and
$N^{i}=N^{i}(t,x^{i})$ is the shift vector.  The spacetime interval between two points
$(t,x^{i})$ and $(t+dt,x^{i}+dx^{i})$ on $\mathcal{M}$ is then
\bea
ds^{2} = -N^{2}dt^{2} + q_{ij}(dx^{i}+N^{i}dt)(dx^{j}+N^{j}dt).
\eea
This is shown in Fig. 4.  In components this gives the metric tensor and its inverse
\bea
g_{\mu\nu} &=&
\left(\begin{array}{cc}
        -N^{2}+N_{i}N^{i} & N_{j}
\\      N_{i} & q_{ij}
\end{array}\right)\\
g^{\mu\nu} &=&
\left(\begin{array}{cc}
        -\frac{1}{N^{2}} & \frac{N^{j}}{N^{2}}
\\      \frac{N^{i}}{N^{2}} & q^{ij}-\frac{N^{i}N^{j}}{N^{2}}
\end{array}\right).
\eea
Here $q^{ij}$ is the inverse of $q_{ij}$.  From the form of the inverse metric above we note that
$q^{ij}$ are not the spatial components of $g^{\mu\nu}$.
\begin{figure}[t]
\centering
\includegraphics[width=2.7in]{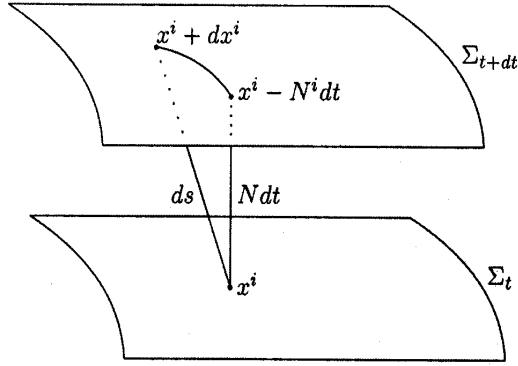}
\caption{The line element expressed in the ADM decomposition.}
\end{figure}

Using the induced metric we can determine the curvature tensor $R^{i}_{\phantom{j}jkl}$ that is
intrinsic to the hypersurfaces $\Sigma_{t}$.  This is determined by parallel transporting tangent
vectors on $\Sigma_{t}$.  However, we can also define an extrinsic curvature to describe how the
hypersurfaces curve with respect to the embedding spacetime.  This is described by the behaviour
of vectors normal to $\Sigma_{t}$.  For $n^{\alpha}=(N,0,0,0)$ a unit normal to $\Sigma_{t}$ in
the lapse and shift decomposition above, this is given by
\bea
K_{ij} = \frac{1}{2N}(\dot{q}_{ij} - ^{(3)}\nabla_{i}N_{j} - ^{(3)}\nabla_{j}N_{i}),
\eea
and with this we find that the Einstein-Hilbert action becomes
\bea
S &=& \frac{1}{16\pi G_{N}}\int dt\int_{\Sigma}d^{3}x\sqrt{q}N(^{(3)}R - 2\Lambda + K_{ij}K^{ij}
      - (q_{ij}K^{ij})^{2})\nonumber\\
       &\phantom{=}&  + \mbox{boundary terms}.
\label{presplit}
\eea
Details can be found in \cite{adm,carlip2}.  The canonical momenta can now be read off from
the action:
\bea
\pi^{ij} &=& \fdiff{S}{\dot{q}_{ij}} = \sqrt{q}(K^{ij} - q^{ij}(q_{kl}K^{kl}))\\
\pi^{0} &=& \fdiff{S}{\dot{N}} = 0\\
\pi^{i} &=& \fdiff{S}{\dot{N}_{i}} = 0.
\eea
There are four primary constraints $\pi^{0}$ and $\pi^{i}$.  The first momentum can be inverted
to give the extrinsic curvature
\bea
K^{ij} = \frac{1}{\sqrt{q}}(\pi^{ij} - q^{ij}(q_{kl}\pi^{kl}))
\eea
and hence the action (\ref{presplit}) becomes
\bea
S &=& \frac{1}{16\pi G_{N}}\int dt\int_{\Sigma}d^{3}x(\pi^{ij}\dot{q}_{ij}
      - N\mathcal{H} - N_{i}\mathcal{H}^{i})\\
\mathcal{H} &=& \frac{1}{\sqrt{q}}(\pi_{ij}\pi^{ij} - (q_{ij}\pi^{ij})^{2})
                - \sqrt{q}(^{(3)}R - 2\Lambda) \approx 0\\
\mathcal{H}^{i} &=& -2 ^{(3)}\nabla_{j}\pi^{ij} \approx 0.
\label{split}
\eea
Here, the Hamiltonian constraint $\mathcal{H}$ and momentum constraints $\mathcal{H}^{i}$ are
secondary constraints, and the lapse and shift are Lagrange multipliers.  The action (\ref{split})
is in the form of a Lagandre transform $\smallint p_{\mu}\dot{q}^{\mu}-H$ for which the term
$\pi^{ij}\dot{q}_{ij}$ plays the role of the term ``$p_{\mu}\dot{q}^{\mu}$''.  Thus we identify
the quantity $H_{ADM}=N\mathcal{H}+N_{i}\mathcal{H}^{i}$ as the gravitational Hamiltonian
(called the ADM Hamiltonian for Arnowitt-Deser-Misner).  Because the Hamiltonian is a linear
combination of constraints it vanishes on-shell (i.e. when the equations of motion are satisfied).
From $H_{ADM}$ we find the equal-time Poisson brackets
\bea
\poisson{q_{ij}(x)}{\pi^{kl}(y)} &=& 8\pi G_{N}
                                   (\delta_{i}^{k}\delta_{j}^{l} + \delta_{i}^{l}\delta_{j}^{k})
                                   \tilde{\delta}^{3}(x-y)\\
\poisson{q_{ij}(x)}{q_{kl}(y)} &=& 0\\
\poisson{\pi^{ij}(x)}{\pi^{kl}(y)} &=& 0,
\eea
and $\smallint d^{3}x\tilde{\delta}^{3}(x-y)f(y)=f(x)$ for any function $f(x)$.

\end{document}